\documentclass[final]{ias2}
\usepackage{bbm}
\usepackage[final]{graphics}
\usepackage{multirow}
\usepackage{array} 

\usepackage{hyperref} 
\usepackage{amsmath}
\usepackage{amsfonts,amsbsy}
\usepackage{amssymb}
\usepackage{bm}

\def\empile#1\over#2{\mathrel{\mathop{\kern 0pt#1}\limits_{#2}}}
\def\bs{\boldsymbol}

\def\TODO#1{}

\def\p{{\boldsymbol p}}

\def\k{{\boldsymbol k}}

\def\x{{\boldsymbol x}}

\def\u{{\boldsymbol u}}
\def\v{{\boldsymbol v}}

\def\colora{}
\def\colorb{}
\def\colorc{}
\def\colord{}

\begin{document}

\markboth{The initial stages of heavy-ion collisions}{Fran\c cois Gelis}

\title{The initial stages of heavy-ion collisions\\in the Color Glass Condensate framework}

\author[ipht]{Fran\c cois Gelis} 
\email{francois.gelis@cea.fr}
\address[ipht]{Institut de Physique Th\'eorique, CEA Saclay, 91191 Gif sur Yvette cedex, France}

\begin{abstract}
  In this short review, we present the description of the early stages
  of a heavy ion collision at high energy in the Color Glass
  Condensate framework.
\end{abstract}

\keywords{Heavy ion collisions, Quantum Chromodynamics, Color Glass Condensate}

\pacs{}
 
\maketitle


\section{Introduction}
\begin{figure}[htbp]
\begin{center}
\resizebox*{9cm}{!}{\includegraphics{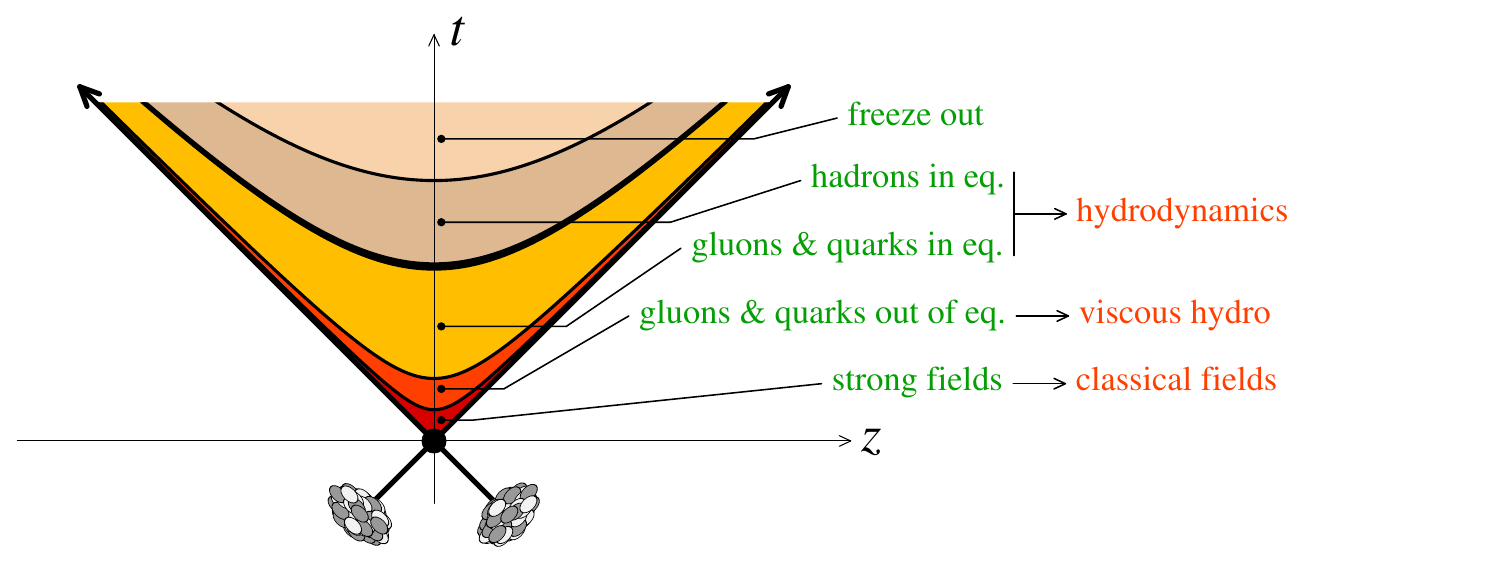}}
\end{center}
\caption{\label{fig:AAstages}Successive stages of a heavy ion collision.}
\end{figure}

Heavy ion collisions at ultra-relativistic energies are used as a way
to study in the laboratory the properties of nuclear matter in extreme
conditions of temperature and density, where it is expected to undergo
a transition to a deconfined state called the quark-gluon
plasma~\cite{KarscC1}. In such a collision, one can distinguish
several stages, illustrated in the figure \ref{fig:AAstages}, from the
initial impact between the two nuclei to the final freeze-out after
which the produced particles stop interacting.
As the matter formed in the collision evolves, it expands and cools
down, which also implies that the strong coupling constant becomes
progressively larger. Therefore, only the first stages (the impact
itself, and shortly afterwards) stand a chance to be amenable to a
description in terms of perturbative Quantum Chromo-Dynamics (QCD).
The bulk evolution of the system during the subsequent stages is well
described by means of transport models such as relativistic
hydrodynamics~\cite{Adamsa3,Adcoxa1,Arsena2,Backa2,HuoviR1,Romat1,Teane1,RomatR1},
that only borrow a few inputs from QCD, in the form of transport
coefficients and initial conditions. The aim of this review is to
discuss our present understanding of the QCD description of the early
stages of heavy ion collisions, up to the stage where the
hydrodynamical description may become applicable.

\section{Gluon saturation, McLerran-Venugopalan model}
\begin{figure}[htbp]
\begin{center}
\resizebox*{6cm}{!}{\includegraphics{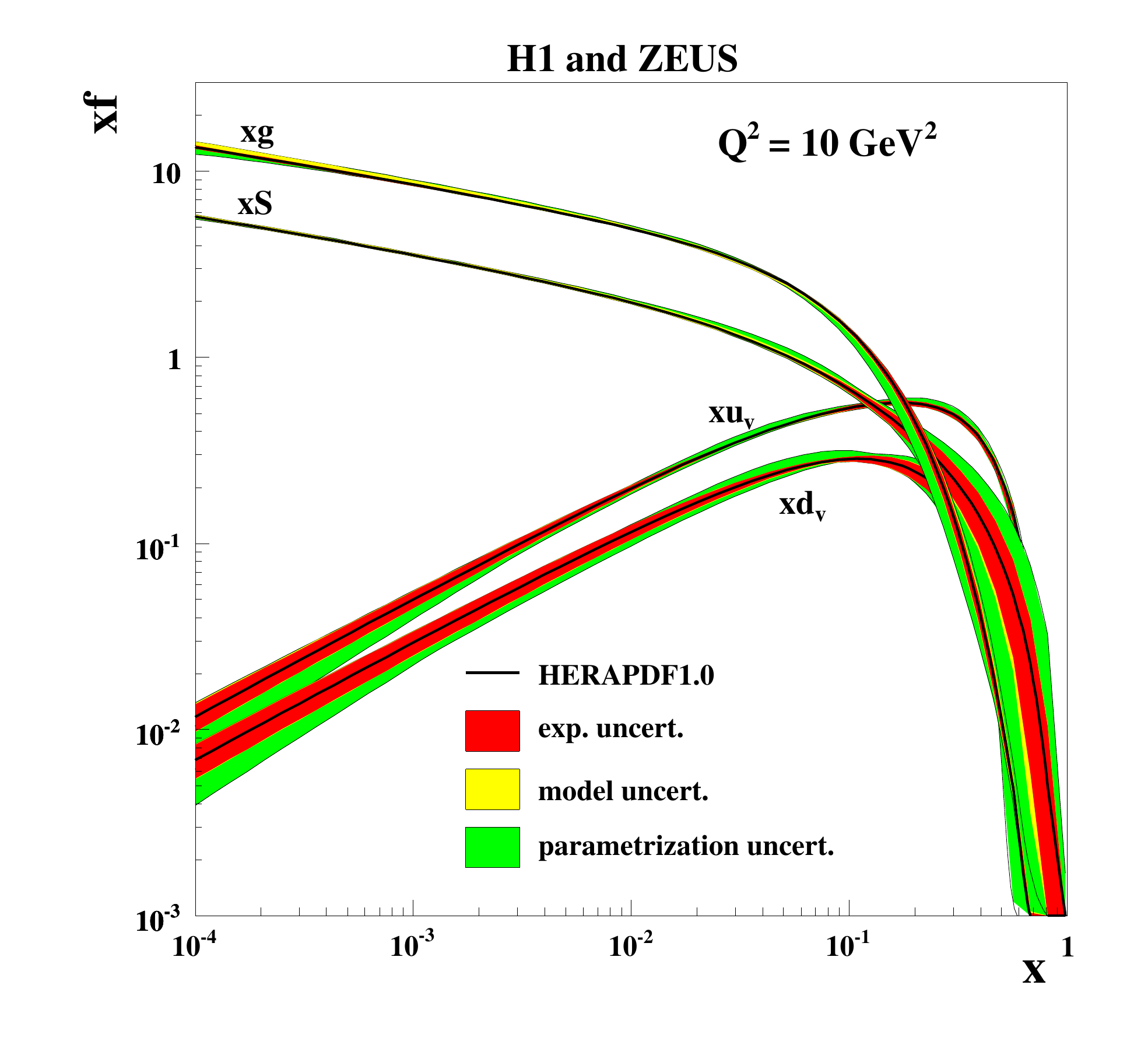}}
\end{center}
\caption{\label{fig:pdf}Parton distributions in a proton. From \cite{Aarona2}.}
\end{figure}

\begin{figure}[htbp]
\begin{center}
\resizebox*{5.5cm}{!}{\includegraphics{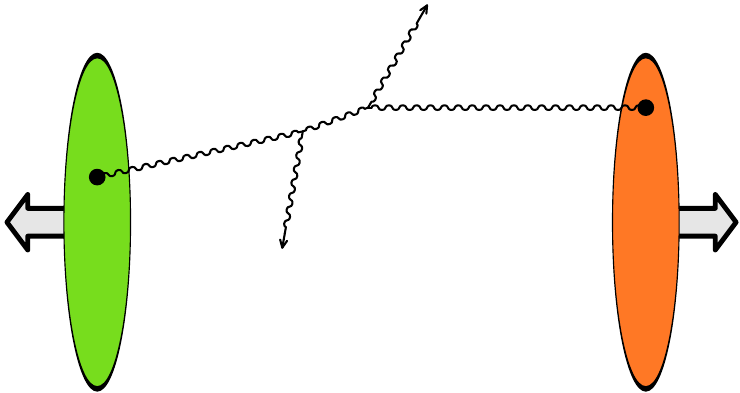}}
\hfill
\resizebox*{5.5cm}{!}{\includegraphics{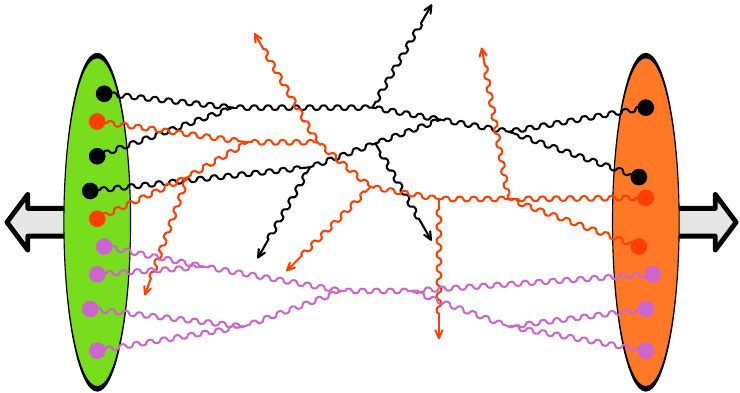}}
\end{center}
\caption{\label{fig:dilute-dense}Typical process in a nucleus-nucleus
  collision. Left: dilute regime. Right: dense regime.}
\end{figure}

In fact, the application of QCD to the study of heavy ion collisions
faces two difficulties. Firstly, it is not at all obvious that the
problem of calculating the bulk properties of the produced quarks and
gluons can be treated by perturbative techniques. Indeed, most of the
partons produced in such a collision are much softer than the
collision energy itself. For a perturbative approach to be possible,
one needs to justify that these partons are produced with a
typical momentum much larger than the QCD non-perturbative scale
$\Lambda_{_{\rm QCD}}$. The second difficulty is that the parton
density in nucleons becomes very large at high energy, as shown in the
figure \ref{fig:pdf} (the longitudinal momentum fraction $x$ is
inversely proportional to the nucleon energy).
When the occupation number in the projectiles approaches the inverse
of the coupling constant, the calculation of transition amplitudes
becomes non-perturbative and cannot be done with the standard
perturbative techniques. The difference between the dilute (left) and
dense (right) regimes is illustrated in the figure \ref{fig:dilute-dense}.
At high density, processes initiated by more than two partons become
important (the power counting relevant for this regime will be
discussed in detail later), and multiple disconnected subprocesses can
occur simultaneously.

\begin{figure}[htbp]
\begin{center}
\resizebox*{6cm}{!}{\includegraphics{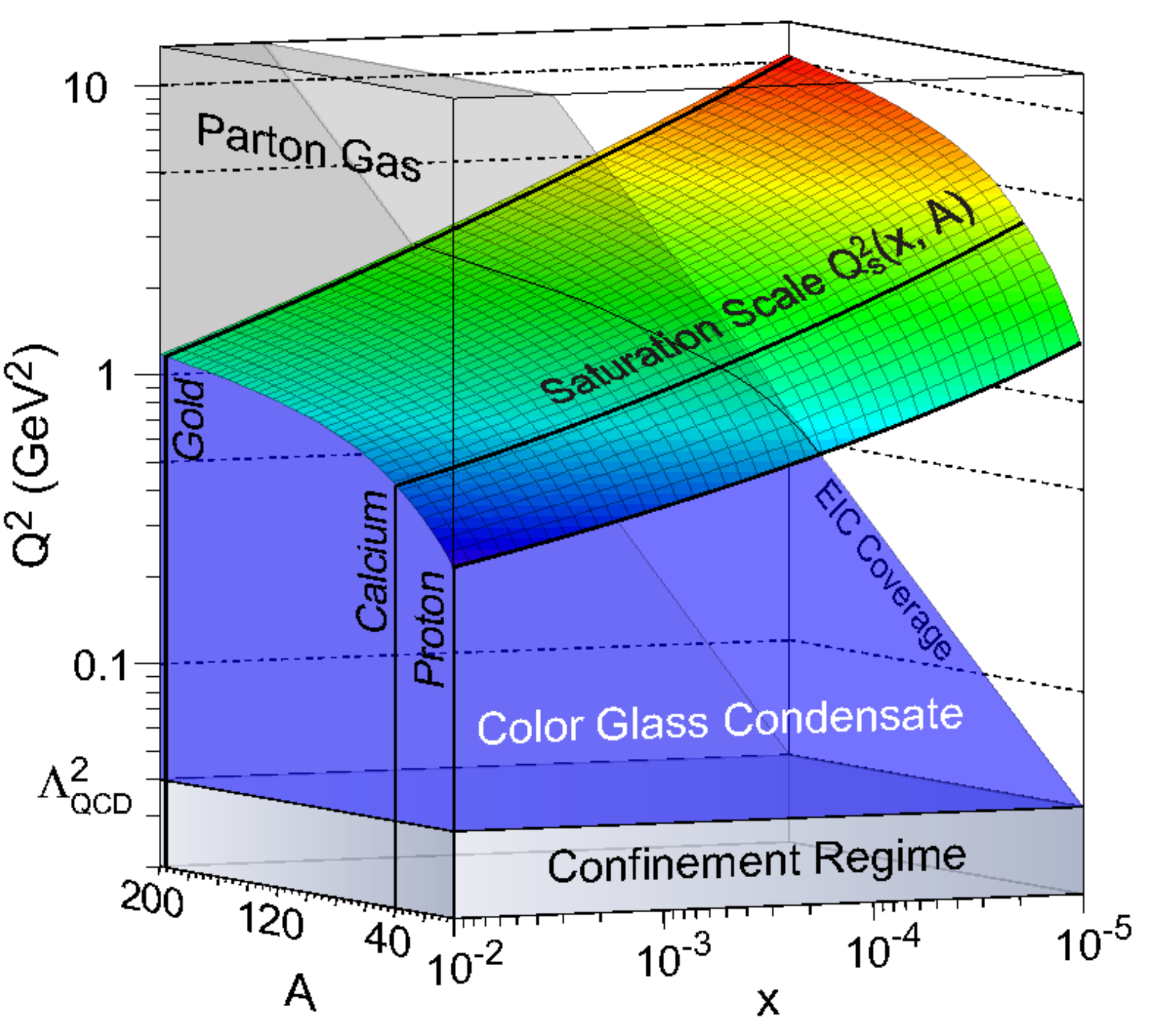}}
\end{center}
\caption{\label{fig:QxA}Saturation domain as a function of $x$ and $A$. From \cite{DeshpEM1}.}
\end{figure}

These two difficulties are in fact intimately related: the non-linear
corrections that become important in the dense regime lead to the
dynamical generation of a characteristic momentum scale --called the
saturation momentum, denoted $Q_s$-- proportional to the gluon
density. The saturation momentum increases with the gluon density, and
therefore with the collision energy, and at sufficiently high energy
it is much larger than $\Lambda_{_{\rm QCD}}$, justifying a weak
coupling treatment. It owes its name to the fact that these non-linear
effects tend to limit --to saturate-- the growth of the gluon
occupation number once it reaches values of order
$\alpha_s^{-1}$. Since the relevant density is the surface density,
i.e. integrated over the collision direction, the saturation momentum
also increases like the radius of the nucleus under consideration,
proportional to the third power of the atomic number, $A^{1/3}$. The
variations of $Q_s$ with $A$ and with the momentum fraction $x$ are
represented in the figure \ref{fig:QxA}.
For lead nuclei ($A\sim 200$) at LHC energy (i.e. $x\sim 10^{-4}$),
the value of $Q_s^2$ is in the range $2-4~$GeV${}^2$, which should be
sufficiently above $\Lambda_{_{\rm QCD}}\approx 200~$MeV to justify a
weak coupling expansion.

The concept of gluon saturation goes back to the early 80's
\cite{GriboLR1,MuellQ1,BlaizM2}. Later on, these ideas were
implemented in the McLerran-Venugopalan model
\cite{McLerV1,McLerV2,McLerV3}, in which the fast partons are
described as classical color sources moving on the light-cone, while
the slower partons (mostly gluons) are described in terms of ordinary
quantum fields. It is thus described by the following effective
Lagrangian,
\begin{equation}
{\cal L}\equiv -\frac{1}{4}F_{\mu\nu}F^{\mu\nu} + J_\mu A^\mu\; ,
\label{eq:L}
\end{equation}
where $J_\mu$ is the color current carried by the fast partons. In the
case of the collision between two hadrons or nuclei, this current is
the sum of two terms, representing color charges moving in opposite
directions. There is an implicit cutoff $\Lambda$ that separates the
fast from the slow partons, depending on their longitudinal momentum.
These color sources are static, because the transverse motion of the
color charges inside the wave function of the projectiles is
considerably slowed down by time dilation, and they appear as frozen
during the short duration of the collision. In a given collision,
$J^\mu$ is a snapshot of the transverse position of the color charges
inside the projectiles. As a consequence, it cannot be known
deterministically, and one can only hope to have a probability
distribution for $J$, usually denoted $W[J]$. Observables should be
calculated for an arbitrary $J$, and then averaged over this
probability distribution.

When this formalism is applied to the study of a collision, the
initial state of the system is the ``vacuum'', since the two
projectiles are entirely encoded in the current $J_\mu$. Thus, to
compute the expectation value of an observable $\widehat{\cal O}$, one
must in fact evaluate $\big<0_{\rm in}\big|\widehat{\cal O}\big|0_{\rm
  in}\big>$ in the quantum field theory described by eq.~(\ref{eq:L}).
The best tool to organize the calculation of this type of matrix
element is the Schwinger-Keldysh formalism~\cite{Schwi1,Keldy1} (that
can also been viewed in this context as a realization of Cutkosky's
cutting rules~\cite{Cutko1,t'HooV1}).

In the saturated regime, the color current $J$ is of order $Q_s^3
g^{-1}$ (the factor $Q_s^3$ just sets the correct dimension). The
inverse power of the coupling, $g^{-1}$, is due to the fact that the
occupation number (quadratic in $J$) is of order $g^{-2}$ in the
saturated regime. This factor deeply alters the power counting when
applying this effective description to the calculation of
observables. In order to see this, consider any connected subgraph of
the diagram shown in the right panel of the figure
\ref{fig:dilute-dense}. In the saturated regime, the order of such a
graph is~\cite{GelisV2,GelisV3}
\begin{equation}
\frac{1}{g^2} \times g^{\rm \#\ of\ produced\ gluons}\times g^{2({\rm \#\ of\ loops})}\; ,
\label{eq:PC}
\end{equation}
which does not depend on the number of times the external source $J$
is inserted into the graph. The loop expansion still corresponds to an
expansion in powers of $g^2$, but each order in this expansion
receives contributions from an infinite set of Feynman graphs that
have a fixed number of loops but differ in how many sources $J$ they
contain.

\begin{figure}[htbp]
\begin{center}
\resizebox*{9cm}{!}{\includegraphics{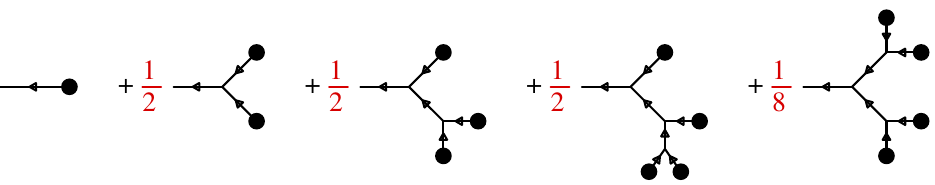}}
\end{center}
\caption{\label{fig:class}The first few terms in the solution of the
  classical equation of motion.}
\end{figure}

\begin{figure}[htbp]
\begin{center}
\resizebox*{4cm}{!}{\includegraphics{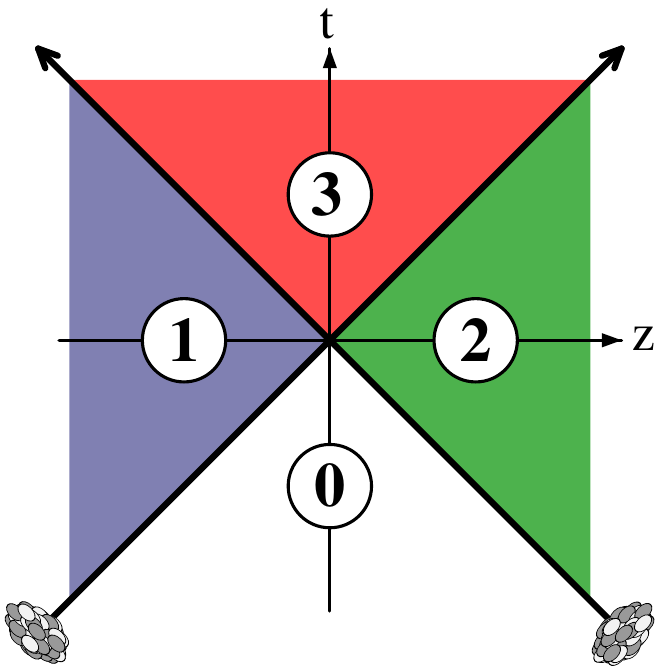}}
\end{center}
\caption{\label{fig:spacetime}Division of space-time in four
  regions. Region 0 and empty. Regions 1 and 2 are causally
  disconnected and describe a single nucleus. Only region 3 contains
  information about the collision.}
\end{figure}

Consider an observable that depends on the gluon field operator,
${\cal O}[\widehat{A}(x)]$ (an example could be the energy momentum
tensor carried by the gluons produced in the collision). At leading
order in $g^2$, one can show that this observable is expressible in
terms of retarded classical solutions of the classical field equations
of motion,
\begin{equation}
\big<0_{\rm in}\big|{\cal O}[\widehat{A}(x)]\big|0_{\rm in}\big>_{_{\rm LO}}
=
{\cal O}[{\cal A}(x)]\; ,
\end{equation}
where
\begin{equation}
\big[{\cal D}_\mu,{\cal F}^{\mu\nu}\big]=J^\nu\quad,\quad \lim_{x^0\to -\infty}{\cal A}^\mu(x)=0\; .
\end{equation}
The solution of the classical Yang-Mills equations sums
the infinite series of tree graphs represented in the figure
\ref{fig:class}, where the black dots represent the external source
$J$ (only cubic vertices are represented, but in Yang-Mills theory
there is also a quartic vertex).
The retarded boundary condition stating that the field must vanish in
the remote past is related to the fact that the initial state is the
vacuum. The derivation of the observable in the Schwinger-Keldysh
formalism provides a more rigorous justification of this boundary
condition. In practical uses in heavy ion collisions, one can solve
analytically the classical Yang-Mills equations up to the forward
light-cone (i.e. in the regions 0, 1 and 2 of the figure
\ref{fig:spacetime}~\cite{Kovch1}), i.e. the hypersurface where the
proper time is $\tau=0^+$~\cite{KovneMW2}.
Beyond this time (i.e. in the region 3 in the figure
\ref{fig:spacetime}), the analytical solution is not known, but it is
straightforward to solve the equations of motion numerically by
discretizing space on a lattice (time remains a continuously varying
variable)~\cite{KrasnV3,KrasnV1,KrasnV2,KrasnNV2,KrasnNV1,Lappi1,Lappi3,KrasnNV3,KrasnNV4,LappiV1}.

\section{Factorization of the distributions of sources, Color Glass Condensate}
The calculation at leading order of the energy-momentum tensor of the
system after the collision already provides some insight on the energy
density released in the collision. However, the LO predictions for the
pressure tensor are more problematic. Immediately after the collision,
the longitudinal pressure is exactly the opposite of the energy
density, because the chromo-electric and magnetic field lines are
parallel to the collision axis~\cite{LappiM1}. At later times, the
longitudinal pressure at LO always remains much smaller than the
transverse pressure~\cite{LappiM1,FukusG1}, which if taken at face
value would cast some doubts about the applicability of hydrodynamics
to describe the bulk evolution of the system.

The hope to find large corrections at next-to-leading order, and
therefore alter this pessimistic conclusion, is the main motivation to
go beyond the leading order. At one loop, one encounters in fact two
types of large corrections. The first one (the second type of large
corrections, related to the Weibel instability, will be discussed in
the next section) is made of contributions that contain logarithms of
the cutoff $\Lambda$ that separates the fast and the slow
partons~\cite{AyalaJMV2}. $\Lambda$ enters in 1-loop corrections as
the upper bound on the longitudinal loop momentum, necessary in order
to avoid double counting the fast modes that are already described as
external sources.  The resummation of these large logarithms, that we
shall describe in this section, promotes the McLerran-Venugopalan
model into a full-fledged effective theory called the Color Glass
Condensate\footnote{For more detailed reviews of the Color Glass
  Condensate, one can consult
  Refs.~\cite{IancuV1,Lappi6,Weige2,GelisIJV1}.}.

The first step is to obtain a formula for the NLO corrections. For the
expectation value of any inclusive observable, the NLO contribution
can be written as follows~\cite{GelisLV3,GelisLV4}
\begin{equation}
\big<0_{\rm in}\big|{\cal O}[\widehat{A}(x)]\big|0_{\rm in}\big>_{_{\rm NLO}}
=
\Bigg[
\int_{\u} {\bs\alpha}(\u){\mathbbm T}_\u
+\frac{1}{2}
\int_{\u\v}
{\bs\Gamma}_2(\u,\v){\mathbbm T}_\u{\mathbbm T}_\v
\Bigg]\,{\cal O}[{\cal A}(x)]\; .
\label{eq:NLO}
\end{equation}
The integrations are over the space-like surface where the initial
conditions for ${\cal A}$ are set, and ${\bs\alpha}$ and ${\bs\Gamma}_2$ are
1- and 2-point functions, evaluated on this surface, that can be
computed in perturbation theory. The operator ${\mathbbm T}_\u$ is the
generator of shifts of the value of the classical field at the point
$\u$ on this surface (roughly speaking, ${\mathbbm T}_\u=\delta/\delta{\cal A}_{\rm init}(\u)$).
\begin{figure}[htbp]
\begin{center}
\resizebox*{4cm}{!}{\includegraphics{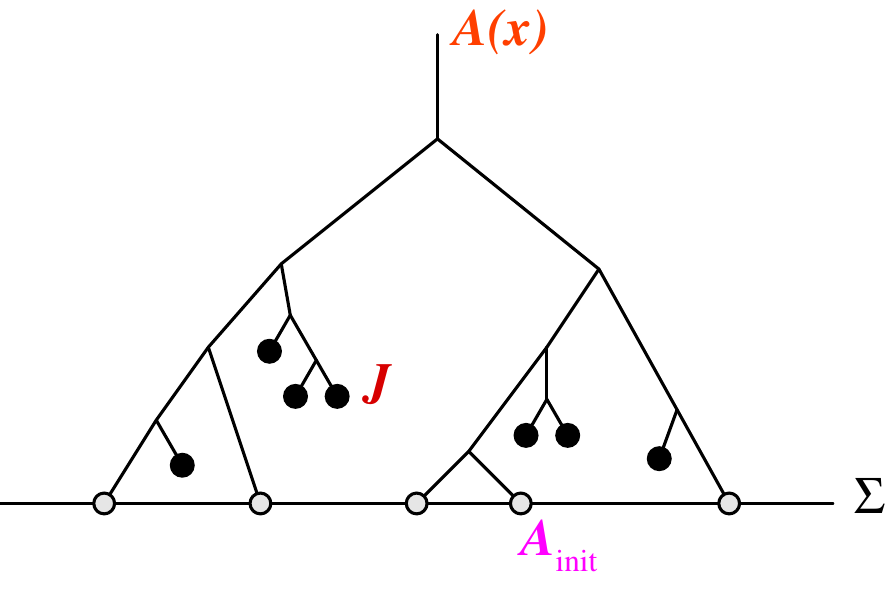}}
\hfill
\resizebox*{4cm}{!}{\includegraphics{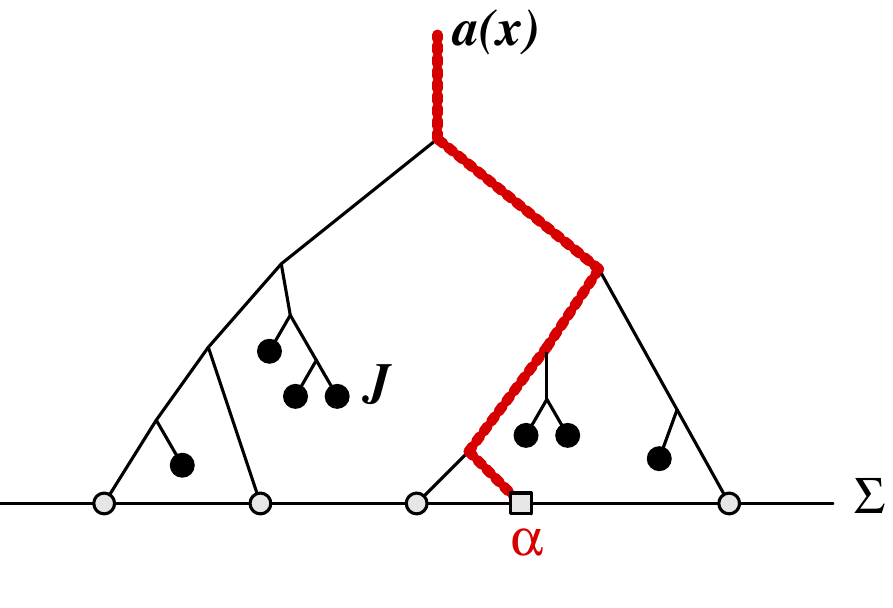}}
\hfill
\resizebox*{4cm}{!}{\includegraphics{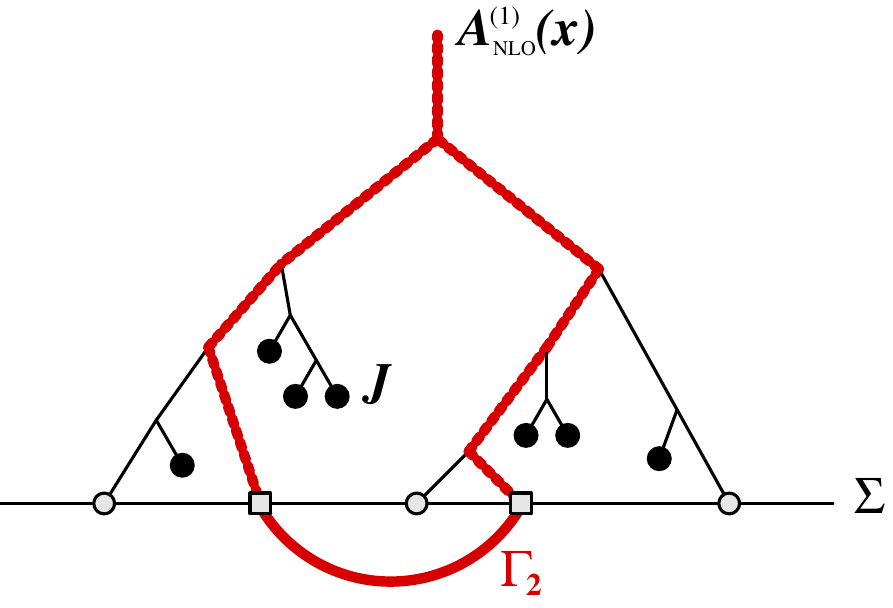}}
\end{center}
\caption{\label{fig:shift-op}Illustration of the action of the shift
  operator ${\mathbbm T}_\u$.}
\end{figure}
One can understand pictorially the action of this operator in the
figure \ref{fig:shift-op}. The left diagram shows that the dependence
of the classical field ${\cal A}(x)$ on its initial condition has a
tree structure. In the middle diagram, we see that by replacing at one
point the initial field ${\cal A}_{\rm init}$ by some quantity
${\bs\alpha}$ (this is what the operator ${\bs\alpha}(\u){\mathbbm
  T}_\u$ does), one generates the retarded propagator that connects
the points $\u$ and $x$, fully dressed by the background classical
field. In the diagram on the right, we see that by acting with
${\bs\Gamma}_2(\u,\v){\mathbbm T}_\u{\mathbbm T}_\v$, one generates a
loop correction to the classical field. These considerations, plus a
little bookkeeping, are the basis of eq.~(\ref{eq:NLO}).

In eq.~(\ref{eq:NLO}), the space-like surface on which the initial
fields are set can be chosen at will, and the left hand side is
independent of this choice. In order to extract the large logarithms
that arise at NLO, it is convenient to choose a surface located just
above the past light-cone (i.e. the wedge made of the lower borders of
the regions 1 and 2 in the figure \ref{fig:spacetime}). With this
choice, one can determine  the logarithms analytically~\cite{GelisLV3,GelisLV4,GelisLV5}:
\begin{equation}
\int_{\u} {\bs\alpha}(\u){\mathbbm T}_\u
+\frac{1}{2}
\int_{\u\v}
{\bs\Gamma}_2(\u,\v){\mathbbm T}_\u{\mathbbm T}_\v
=
\log(\Lambda)\Big[{\cal H}_1+{\cal H}_2\Big] + \mbox{terms w/o logs}
\label{eq:logs}
\end{equation}
where ${\cal H}_1$ and ${\cal H}_2$ are operators known as the JIMWLK
Hamiltonians~\cite{Balit1,JalilKMW1,JalilKLW1,JalilKLW2,JalilKLW3,JalilKLW4,IancuLM1,IancuLM2,FerreILM1}
for the two nuclei. A remarkable feature of eq.~(\ref{eq:logs}) is
that the coefficient of the logarithms does not contain terms mixing
the sources of the two nuclei. Instead, it is nicely arranged as the
sum of two terms, one for each nucleus. It is this absence of mixing
that allows these logarithms to be absorbed into separate
distributions of sources for the nucleus 1 and the nucleus 2,
respectively. Schematically, the leading log terms $(\alpha_s
\log\Lambda)^n$ (i.e. at each order in $\alpha_s$, the terms that have
the maximal number of logarithms) can be resummed in formulas such as
\begin{equation}
\big<0_{\rm in}\big|{\cal O}[\widehat{A}(x)]\big|0_{\rm in}\big>_{_{\rm Leading\ Log}}
=
\int [DJ_1\,DJ_2]\;W[J_1]\,W[J_2]\,{\cal O}[{\cal A}_{_{J_{1,2}}}(x)]\; ,
\end{equation}
where ${\cal A}_{_{J_{1,2}}}$ is the classical solution of the
Yang-Mills equations with sources $J_{1,2}$, and the $W[J_{1,2}]$ are
the probability distributions for each projectile. The latter obey the
JIMWLK
equation
that drives their rapidity evolution~:
\begin{equation}
\frac{\partial W}{\partial Y}={\cal H}\,W\; .
\end{equation}
\begin{figure}[htbp]
\begin{center}
\resizebox*{6cm}{!}{\includegraphics{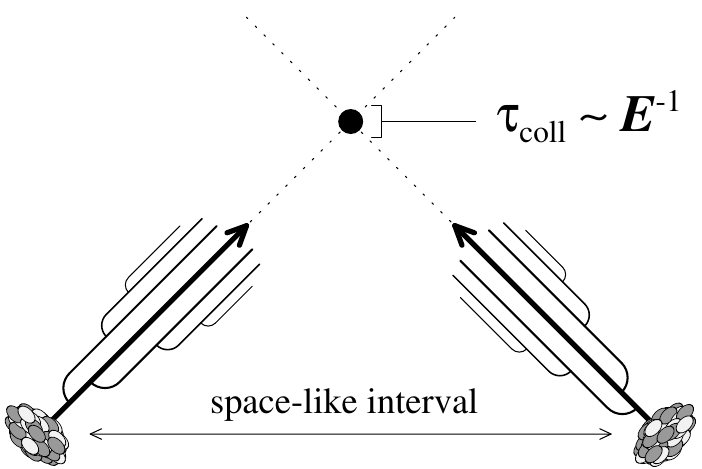}}
\end{center}
\caption{\label{fig:causality}Causality argument for the factorization of the logarithms.}
\end{figure}
The possibility to factorize these logarithms into distributions that
describe the color content of the two projectiles can be understood
qualitatively from a simple causality argument, illustrated in the
figure \ref{fig:causality}. The logarithms come from soft gluon
radiation, which can only occur over long time-scales, and not during
the collision itself, whose duration is very short (inversely
proportional to the collision energy). For this reason, these gluons
must be produced (long) before the collision, at a time when the two
nuclei where not yet in causal contact. This explains why the
coefficients of the logarithms cannot mix the sources of the two
projectiles.

\section{Instabilities, Resummation}
The terms that are logarithmic in the cutoff $\Lambda$ are not the
only large corrections one should worry about at NLO. In order to see
that, it is useful to have in mind the following representation for
the 2-point function ${\bs\Gamma}_2(\u,\v)$ that enters in eq.~(\ref{eq:NLO})~:
\begin{equation}
{\bs\Gamma}_2(\u,\v)
=
\int\frac{d^3\k}{(2\pi)^3 2k}\;
{\bs\alpha}_\k(\u){\bs\alpha}_\k^*(\v)\; ,
\label{eq:modes}
\end{equation}
where ${\bs\alpha}_\k(x)$ obeys the linearized classical equation of
motion around the classical solution ${\cal A}$, and whose initial
condition in the remote past is a plane wave of momentum $\k$. In
other words, the ${\bs\alpha}_\k$ form a complete basis for the vector
space of linear perturbations around the classical solution ${\cal
  A}$.  Note that the modes that give the logarithms of $\Lambda$ are
those that are independent of the rapidity $\eta$.

\begin{figure}[htbp]
\begin{center}
\resizebox*{6.2cm}{!}{\rotatebox{-90}{\includegraphics{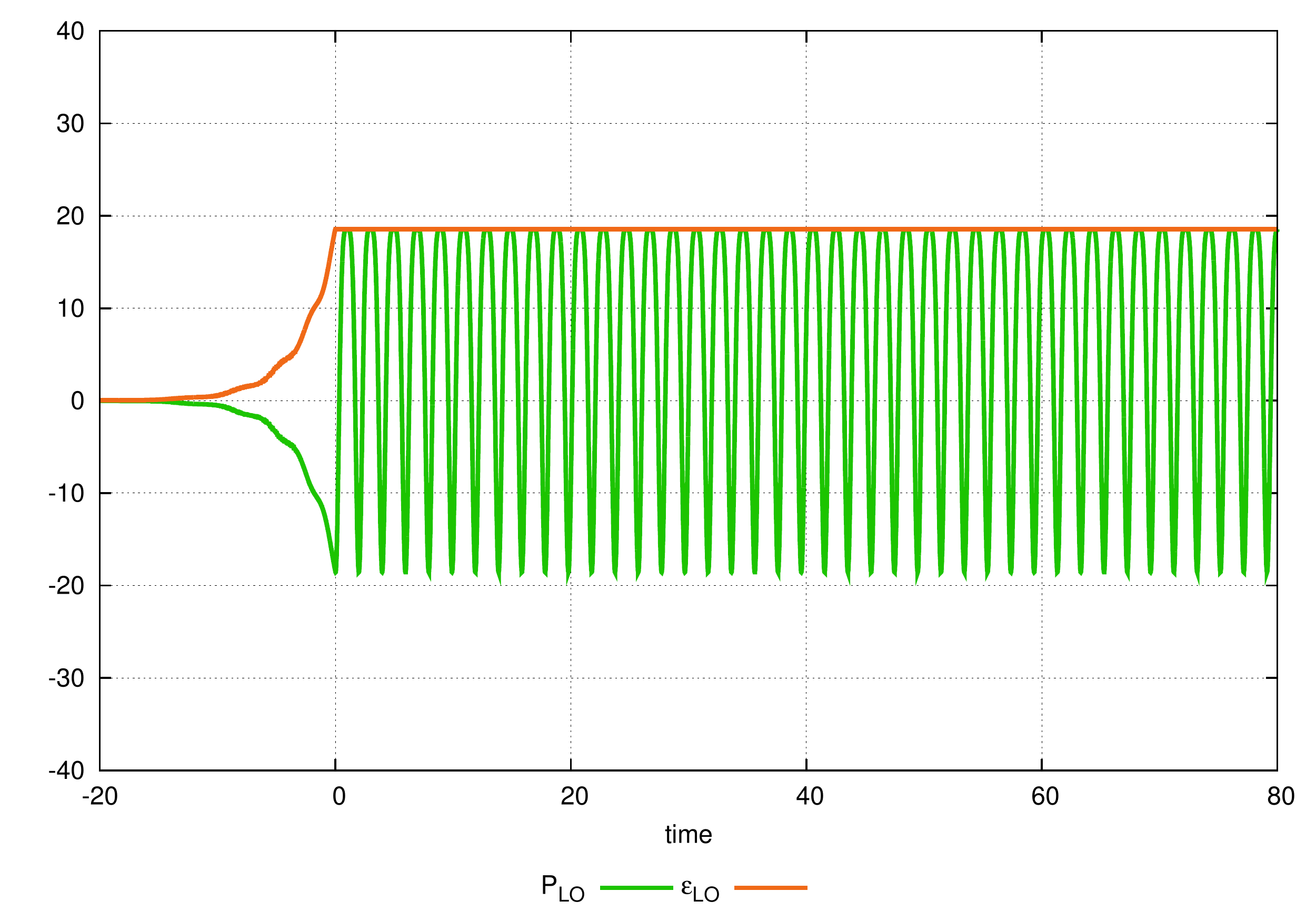}}}
\hfill
\resizebox*{6.2cm}{!}{\rotatebox{-90}{\includegraphics{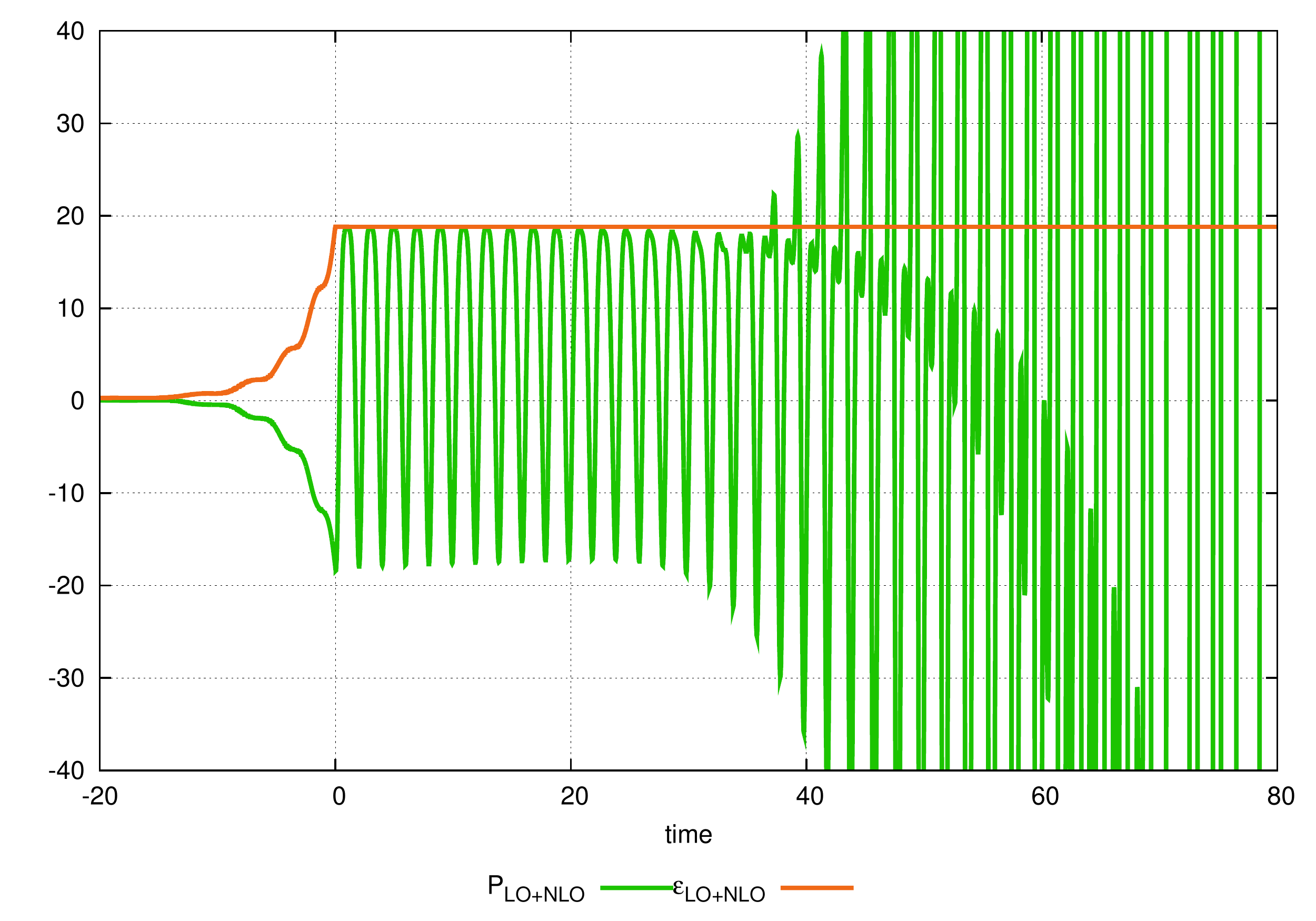}}}
\end{center}
\caption{\label{fig:phi4-insta}Example of secular divergence due to
  instabilities in the classical equations of motion, for the pressure
  in a $\phi^4$ scalar field theory~\cite{DusliEGV1}.}
\end{figure}

As we shall see, large corrections to the observable can also arise
from some of the modes that depend on rapidity. From the definition of
the shift operator ${\mathbbm T}_\u$, we also know that
\begin{equation*}
\left[\int_\u {\bs\alpha}_\k(\u){\mathbbm T}_\u\right]\,{\cal A}(x) = {\bs\alpha}_\k(x)\; .
\end{equation*}
The loop expansion of observables is based on the assumption that the
mode functions ${\bs\alpha}_\k(x)$ remain small at all times.  If for
some $\k$, the perturbation ${\bs\alpha}_\k(x)$ becomes as large as
$g^{-1}$, then the NLO corrections can be as large as the LO
result. It turns out that this is the case for Yang-Mills equations,
that are known to exhibit chaotic behavior at the classical
level~\cite{RomatV1,RomatV2,RomatV3,FukusG1,BiroGMT1,HeinzHLMM1,BolteMS1,FujiiI1,FujiiII1,KunihMOST1}. The
fact that their solutions are exponentially sensitive to the initial
conditions means that the linearized equations of motion have some
mode functions that grow exponentially with time. In fact, these
instabilities of the solutions of the classical equations of motion
are closely related to the Weibel instability in anisotropic
Yang-Mills
plasmas~\cite{Mrowc3,Mrowc4,RebhaRS1,RebhaRS2,MrowcRS1,RomatS1,RomatS2,RebhaS1,RebhaSA1,ArnolLM1,ArnolLMY1,ArnolM1,ArnolM2,ArnolM3,ArnolMY4,DumitNS1,BodekR1,BergeGSS1,BergeSS2,KurkeM1,KurkeM2,AttemRS1}. If
a mode function ${\bs\alpha}_\k$ grows exponentially as $\exp(\mu_\k
t)$, then it gives a term that grows like $\exp(2\mu_\k t)$ in
the NLO correction to the observable\footnote{Note that the term in
  ${\bs\alpha}(\u){\mathbbm T}_\u$ in eq.~(\ref{eq:NLO}) cannot lead
  to a similar problem. Indeed, the 1-point function ${\bs\alpha}(\u)$
  is rapidity independent, and boost invariant mode functions are
  stable.}. One can see the effect of these instabilities in the
calculation of the pressure in a $\phi^4$ scalar field theory (that
also has unstable classical solutions for certain $\k$-modes) in the
figure \ref{fig:phi4-insta}.
The NLO correction is a very small correction at the beginning of the
time evolution, but it grows exponentially with time. At late times,
the loop expansion breaks down and one should seek a resummation that
will collect and resum these large corrections. The first step in
doing this is to improve the power counting in order to keep track of
the growth rate of each contribution.
\begin{figure}[htbp]
\begin{center}
\resizebox*{4cm}{!}{\includegraphics{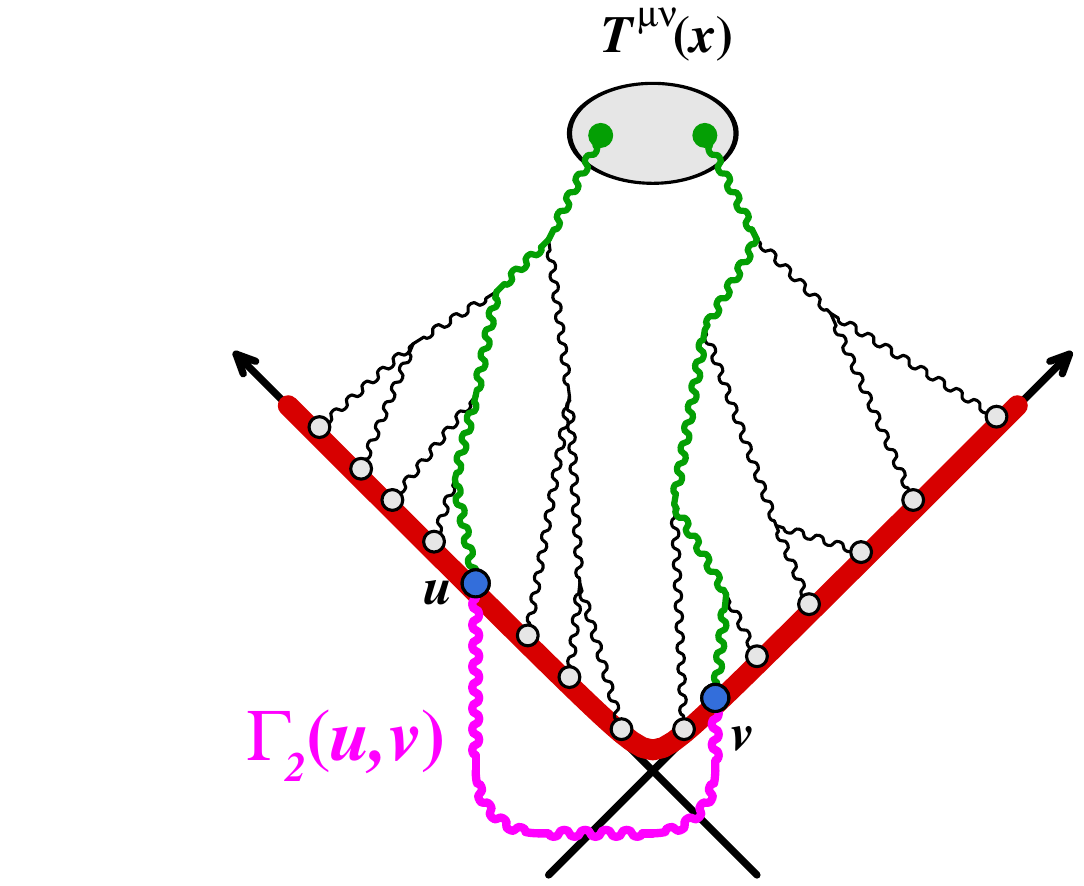}}
\hfill
\resizebox*{4cm}{!}{\includegraphics{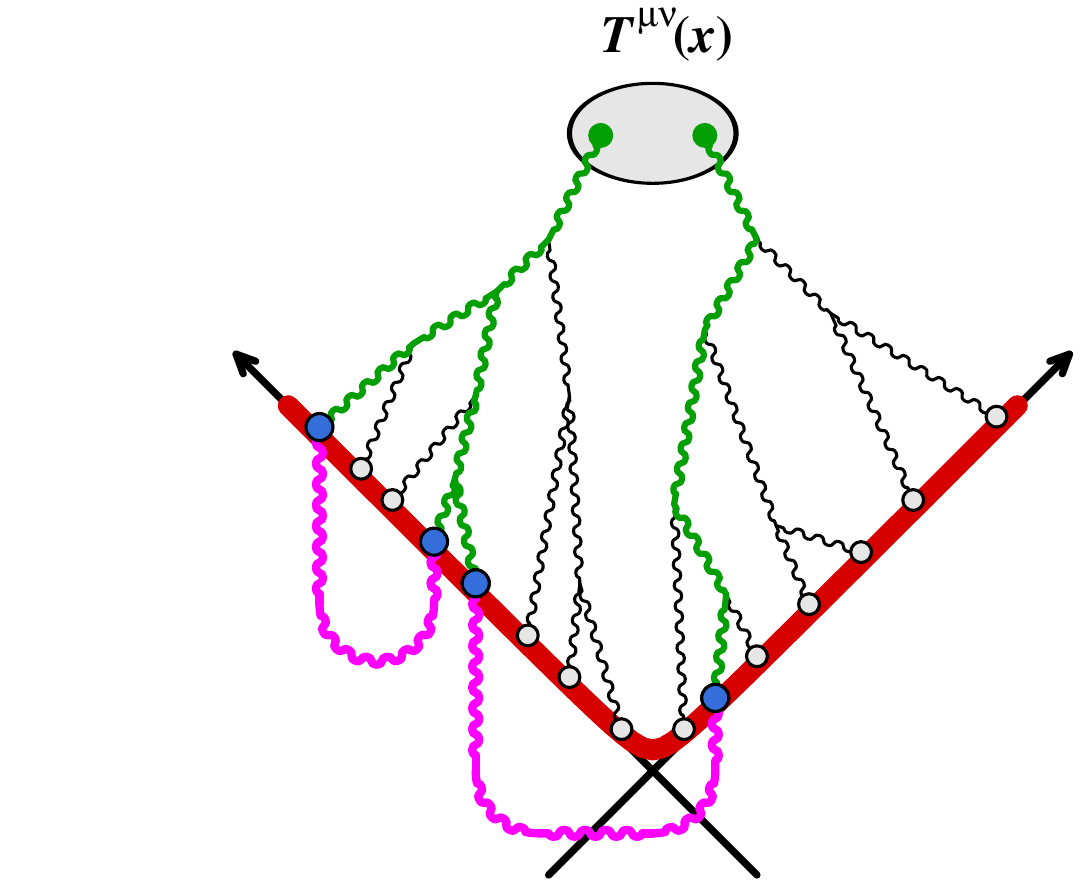}}
\hfill
\resizebox*{4cm}{!}{\includegraphics{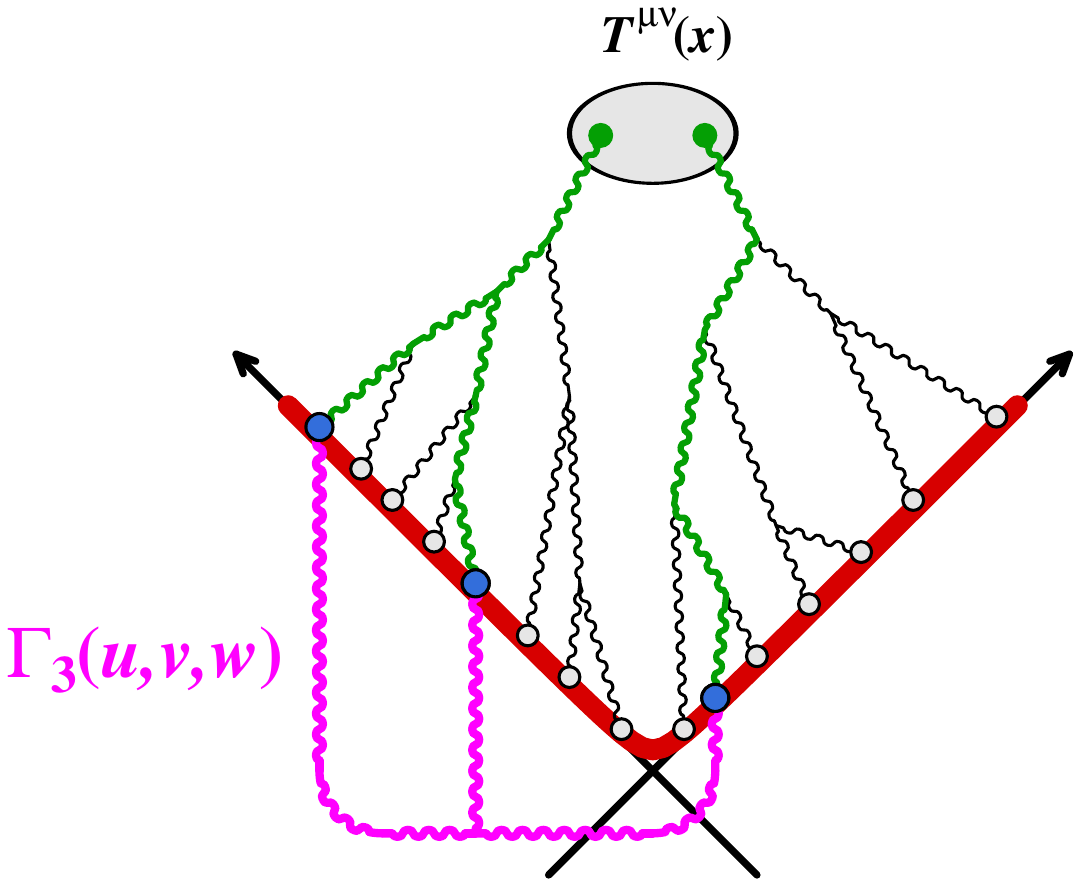}}
\end{center}
\caption{\label{fig:resum}Loop corrections to the energy-momentum tensor.}
\end{figure}
How to do this is illustrated in the figure \ref{fig:resum}. The
standard power counting (see the eq.~(\ref{eq:PC})) indicates that one
gets a power of $g^2$ for each loop. From the above discussion, one
may also get an exponentially growing factor for each operator
${\mathbbm T}$. From the examples of the figure \ref{fig:resum}, we
see that diagrams with $n$ loops can have up to $2n$ insertions of the
operator ${\mathbbm T}$, and that this maximum is reached only when
the loops are independent below the surface used as initial Cauchy
surface. From this observation, it appears that one can resum to all
loop orders the contributions that have the fastest growth by
exponentiating the term in ${\mathbbm T}_\u{\mathbbm T}_\v$ in
eq.~(\ref{eq:NLO})~:
\begin{equation}
\big<0_{\rm in}\big|{\cal O}[\widehat{A}(x)]\big|0_{\rm in}\big>_{_{\rm resummed}}
=
\exp\Bigg[
\frac{1}{2}
\int_{\u\v}
{\bs\Gamma}_2(\u,\v){\mathbbm T}_\u{\mathbbm T}_\v
\Bigg]\,{\cal O}[{\cal A}(x)]\; .
\label{eq:resummed}
\end{equation}
(At this point, we assume that the leading logs in the cutoff
$\Lambda$ have already been taken care of. Not double counting these
contributions is ensured by excluding the rapidity independent
modes in ${\bs\Gamma}_2(\u,\v)$ in eq.~(\ref{eq:resummed}).) The right hand
side of this equation is not easy to evaluate in this form, but it is
equivalent to letting the initial condition of the classical field
${\cal A}$ fluctuate with a Gaussian distribution of variance ${\bs
  G}(\u,\v)$.  More precisely, we have
\begin{eqnarray}
&&\big<0_{\rm in}\big|{\cal O}[\widehat{A}(x)]\big|0_{\rm in}\big>_{_{\rm resummed}}
=\nonumber\\
&&\!\!\!\!\!\!\!\!\!\!\!\!=
\int [D{\bs\chi}(\u)]\;
\exp\Bigg[
-\frac{1}{2}
\int_{\u\v}
{\bs\Gamma}_2^{-1}(\u,\v){\bs\chi}(\u){\bs\chi}(\v)
\Bigg]\,{\cal O}[{\cal A}[{\cal A}_{\rm init}+{\bs\chi}](x)]\; ,
\end{eqnarray}
where the notation ${\cal A}[{\cal A}_{\rm init}+{\bs\chi}](x)$
denotes the value at the point $x$ of the classical solution whose
initial condition is ${\cal A}_{\rm init}+{\bs\chi}$ (${\cal A}_{\rm
  init}$ being the initial condition of the classical field at LO).
This alternative form of the resummed observable can be evaluated by
performing a Monte-Carlo average of classical solutions that have
Gaussian distributed initial conditions. It is important to note that
the variance ${\bs\Gamma}_2(\u,\v)$ of the fluctuating initial
conditions is not arbitrary: it is the same function that
enters in the representation (\ref{eq:NLO}) of the NLO result. This
means that one should first perform a 1-loop calculation in order to
obtain it.  This Gaussian average has also been derived  in different
approaches and contexts~\cite{PolarS1,Son1,KhlebT1,MichaT1,FukusGM1}.

Given its origin as the exponentiation of the 1-loop result, this
resummation contains by construction the complete LO and NLO
contributions (for the latter, this is true only if one uses the
variance ${\bs G}(\u,\v)$ given by the 1-loop calculation), and a
subset of all the higher orders. The possibility to include all NLO
effects solely by altering the initial conditions while keeping the
time evolution classical is an example of the general property that
quantum corrections to the time evolution arise only at NNLO and
beyond. Order $\hbar$ corrections in the initial state come from the
uncertainty principle: the fields and their conjugate momenta cannot
be known simultaneously with absolute accuracy, and their distribution
in phase-space must have a support of area at least $\hbar$ for each
mode.

It is easy to see why this resummation cures the problem of secular
divergences that plagues fixed loop order calculations. Indeed, since
the resummation has promoted linearized fluctuations to fields that
obey the full non-linear equation of motion, and since the potential
is bounded from below (for a theory that has a well defined ground
state), the fields that enter in the evaluation of the resummed observable
cannot run away and generate large contributions.

\begin{figure}[htbp]
\begin{center}
\resizebox*{3cm}{!}{\includegraphics{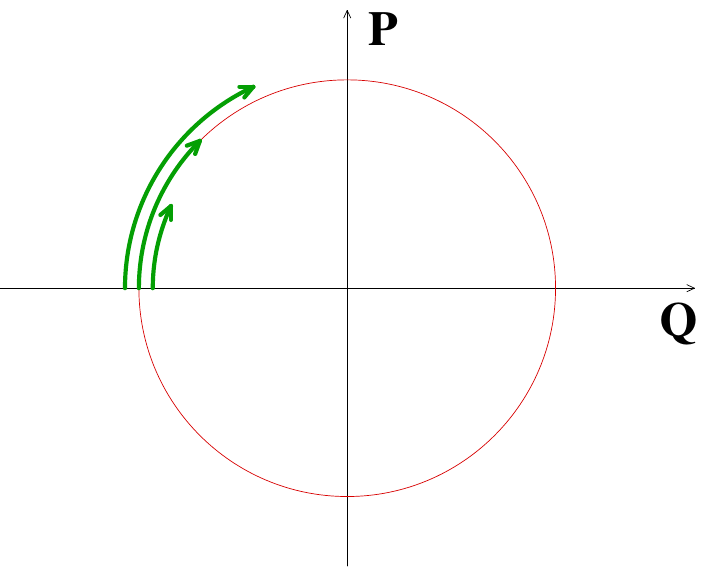}}
\hfill
\resizebox*{3cm}{!}{\includegraphics{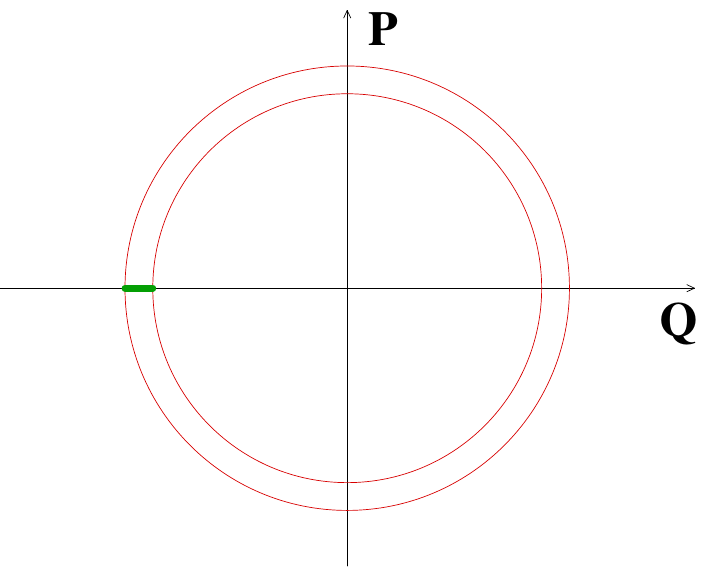}}
\hfill
\resizebox*{3cm}{!}{\includegraphics{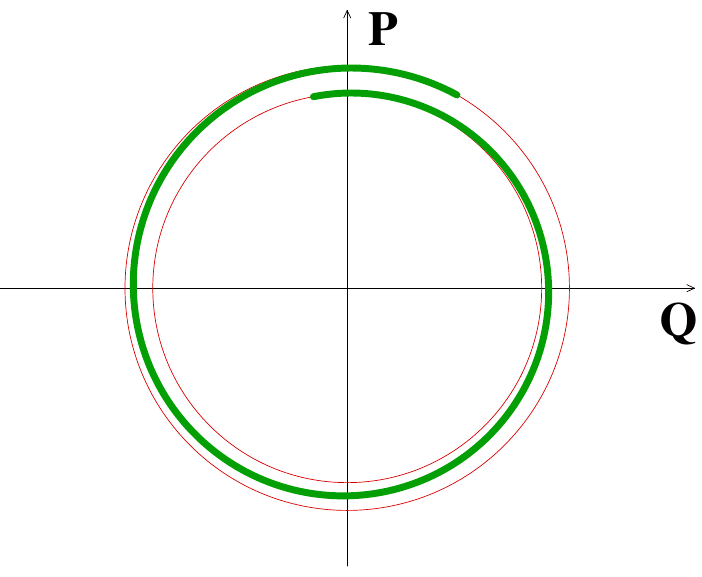}}
\hfill
\resizebox*{3cm}{!}{\includegraphics{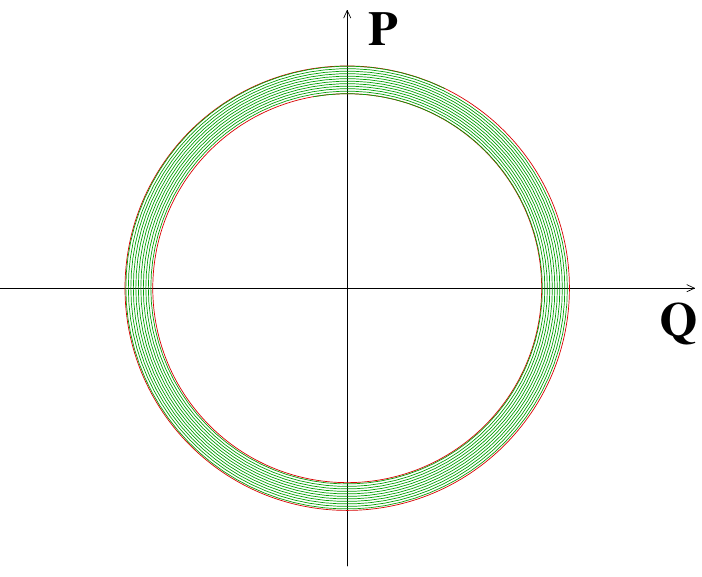}}
\end{center}
\caption{\label{fig:decoherence}Illustration of the decoherence that occurs with non-linear interactions.}
\end{figure}
In the presence of non-linear interactions, one can foresee without
any extensive computation that the resummed result may evolve towards
some form of statistical equilibrium state. This is illustrated in the
figure \ref{fig:decoherence}, for a single non-harmonic oscillator
mode. The left figure shows that the oscillation frequency depends on
the amplitude of the oscillations for such an oscillator. The 2nd, 3rd
and 4th figures (from left to right) then illustrate the time
evolution of an initially narrow Gaussian distribution in
phase-space. The non-harmonicity will cause this distribution to
stretch around the classical orbits, since the outer points rotate at
a frequency that differs from that of the inner points. Eventually,
the distribution covers uniformly the area allowed by energy
conservation, which one may view as a micro-canonical equilibrium
distribution (i.e. a distribution for which all the micro-states
allowed by energy conservation are equally likely).

In order to apply this resummation to heavy ion collisions, we need to
determine the variance ${\bs\Gamma}_2(\u,\v)$ of the spectrum of
fluctuations.  From this spectrum, one will then generate random
initial conditions and solve numerically the classical Yang-Mills
equations. For practical reasons, the numerical resolution of the
equations of motion is started after the collision, at a small but
positive proper time. Because of this, it is necessary to have
analytical expressions for the initial background field and for the
variance of the fluctuating part. The background field at $Q_s\tau\ll
1$ is already known analytically~\cite{KovneMW2}.
\begin{figure}[htbp]
\begin{center}
\resizebox*{4cm}{!}{\includegraphics{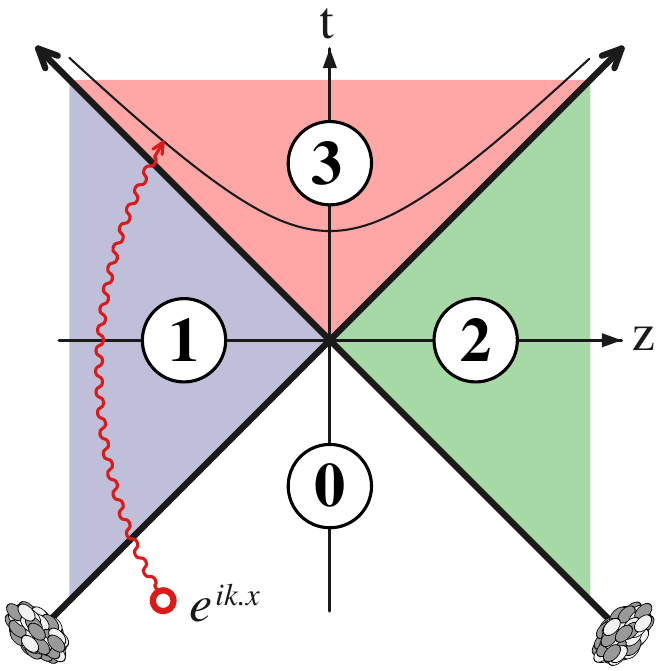}}
\end{center}
\caption{\label{fig:fluct}Evolution of a perturbation on top of the classical background field.}
\end{figure}
In order to calculate the variance, we use the representation of
eq.~(\ref{eq:modes}), which reduces the problem to solving the
linearized equation of motion for small perturbations to the classical
background field, with a plane wave initial condition at $x^0=-\infty$
(i.e. a mode with fixed initial momentum, color, and polarization).
This calculation is illustrated in the figure \ref{fig:fluct}. In
order to be used as initial conditions for the Yang-Mills equations,
we need these mode functions in the same gauge as the one used when
solving numerically the equations, usually the Fock-Schwinger gauge
${\cal A}^\tau=0$. The propagation of the perturbation over the region
0 is of course trivial, since the background field is zero there. The
crossings of the light-cones (where the sources of the two nuclei are
located) are a bit more delicate, because the background field has an
infinite field strength there -- this produces a finite jump of the
fluctuation while crossing the light-cones. Finally, the propagation
in the regions 1 and 2 are rather simple, because the background field
is a pure gauge in these regions. On the proper time surface located
at $Q_s\tau\ll 1$, the perturbation of momentum $\k_\perp,\nu$ ($\nu$
is the Fourier conjugate of the rapidity $\eta$), color $c$ and
polarization $\lambda$ reads~\cite{EpelbG2}
\begin{align}
&{\bs\alpha}^{\colorc i}_{\nu\k_\perp c \lambda}= \beta^{+{\colorc i}}_{\nu\k_\perp c \lambda}+\beta^{-{\colorc i}}_{\nu\k_\perp c \lambda}\;,\,
&{\bs e}^{\colorc i}_{\nu\k_\perp c \lambda}= -i{\colora\nu}\Big(\beta^{+{\colorc i}}_{\nu\k_\perp c \lambda}-\beta^{-{\colorc i}}_{\nu\k_\perp c \lambda}\Big)\nonumber\\
&{\bs\alpha}^{\colorc\eta}_{\nu\k_\perp c \lambda}= {\cal D}^{\colorc i}
\Big(\frac{\beta^{+{\colorc i}}_{\nu\k_\perp c \lambda}}{2+i{\colora\nu}}-\frac{\beta^{-{\colorc i}}_{\nu\k_\perp c \lambda}}{2-i{\colora\nu}}\Big)\;,\,
&{\bs e}^{\colorc\eta}_{\nu\k_\perp c \lambda}=-{\cal D}^{\colorc i}\Big(\beta^{+{\colorc i}}_{\nu\k_\perp c \lambda}-\beta^{-{\colorc i}}_{\nu\k_\perp c \lambda}\Big)\, ,
\end{align}
(the ${\bs e}$'s are the electrical fields associated to the ${\bs\alpha}$'s)
where we denote
\begin{eqnarray}
\beta^{+{\colorc i}}_{\nu\k_\perp c \lambda}&\equiv&
e^{\frac{\pi{\colora\nu}}{2}}\Gamma(-i{\colora\nu}) e^{i{\colora\nu}\eta}\,
{\colorb{\cal U}_1^\dagger(\x_\perp)} 
\smash{\int\limits_{\p_\perp}} e^{i\p_\perp\cdot\x_\perp}\,
{\colorb\widetilde{\cal U}_1(\p_\perp+{\colora\k_\perp})}
\nonumber\\
&&\qquad\qquad\qquad\qquad\qquad\quad\times
\left(\frac{p_\perp^2\tau}{2{\colora k_\perp}}\right)^{i{\colora\nu}}
\!
\Big(\delta^{{\colorc i}{\colord j}}-2\frac{p_\perp^{{\colorc i}} p_\perp^{{\colord j}}}{p_\perp^2}\Big){\colord \epsilon^j_\lambda}\; ,
\end{eqnarray}
and
\begin{eqnarray}
\beta^{-{\colorc i}}_{\nu\k_\perp c \lambda}&\equiv&
e^{-\frac{\pi{\colora\nu}}{2}}\Gamma(i{\colora\nu}) e^{i{\colora\nu}\eta}\,
{\colorb{\cal U}_2^\dagger(\x_\perp)} 
\smash{\int\limits_{\p_\perp}} e^{i\p_\perp\cdot\x_\perp}\,
{\colorb\widetilde{\cal U}_2(\p_\perp+{\colora\k_\perp})}
\nonumber\\
&&\qquad\qquad\qquad\qquad\qquad\quad\times
\left(\frac{p_\perp^2\tau}{2{\colora k_\perp}}\right)^{-i{\colora\nu}}
\!\!
\Big(\delta^{{\colorc i}{\colord j}}-2\frac{p_\perp^{{\colorc i}} p_\perp^{{\colord j}}}{p_\perp^2}\Big){\colord \epsilon^j_\lambda}\; .
\end{eqnarray}
A generic initial condition is the sum of the background field at the
initial time $\tau_0$, and of all the mode functions
${\bs\alpha}_{\nu\k_\perp c \lambda}$, weighted by random Gaussian
distributed complex numbers $c_{\nu\k_\perp c \lambda}$, normalized so that
\begin{equation}
\left<c_{\nu\k_\perp c \lambda}\,c^*_{\nu'\k_\perp' c' \lambda'}\right>=
\frac{1}{2}\delta(\nu-\nu')\delta(\k_\perp-\k_\perp')\delta_{cc'}\delta_{\lambda\lambda'}\; .
\end{equation}

\section{Evolution after the collision}
Thanks to these mode functions, one can implement numerically the
resummation described in the previous section in the case of heavy ion
collisions.  At LO, the chromo-electric and chromo-magnetic fields are
parallel to the collision axis when $Q_s\tau\ll 1$ and the
corresponding longitudinal pressure is negative, opposite to the
energy density, while the transverse pressure equals the energy
density. At later time, the longitudinal pressure increases near zero,
but remains negligible compared to the transverse pressure at all
times.
\begin{figure}[htbp]
\begin{center}
\resizebox*{6cm}{!}{\includegraphics{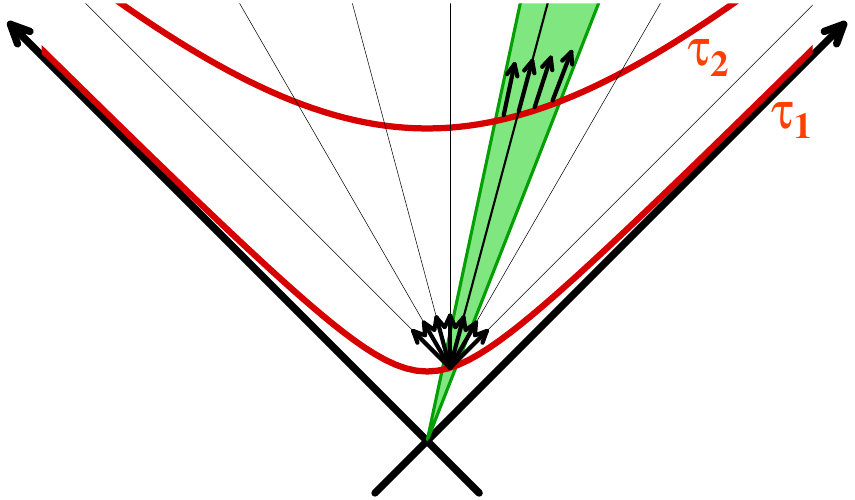}}
\end{center}
\caption{\label{fig:expansion}Evolution of the distribution of
  longitudinal momenta in the comoving frame for a free streaming
  system.}
\end{figure}
This can be understood if the system is nearly free streaming, as
illustrated in the figure \ref{fig:expansion}. If one starts from an
arbitrary distribution of momenta at the time $\tau_1$, and let the
particles evolve freely to the time $\tau_2$, there will be a
segregation of the local momentum distributions, such that only
particles with momentum rapidity $y\approx\eta$ stay at the
rapidity $\eta$. This means that the longitudinal momenta in the
comoving frame are nearly zero, and the longitudinal pressure is
very small.

One of the main physical issues is therefore whether the instabilities
that one is resumming by including Gaussian fluctuations of the
initial fields can efficiently reshuffle the momenta and compete
against the longitudinal expansion of the system, in order to generate
a sizable longitudinal pressure.

\begin{figure}[htbp]
\begin{center}
\resizebox*{7.5cm}{!}{\includegraphics{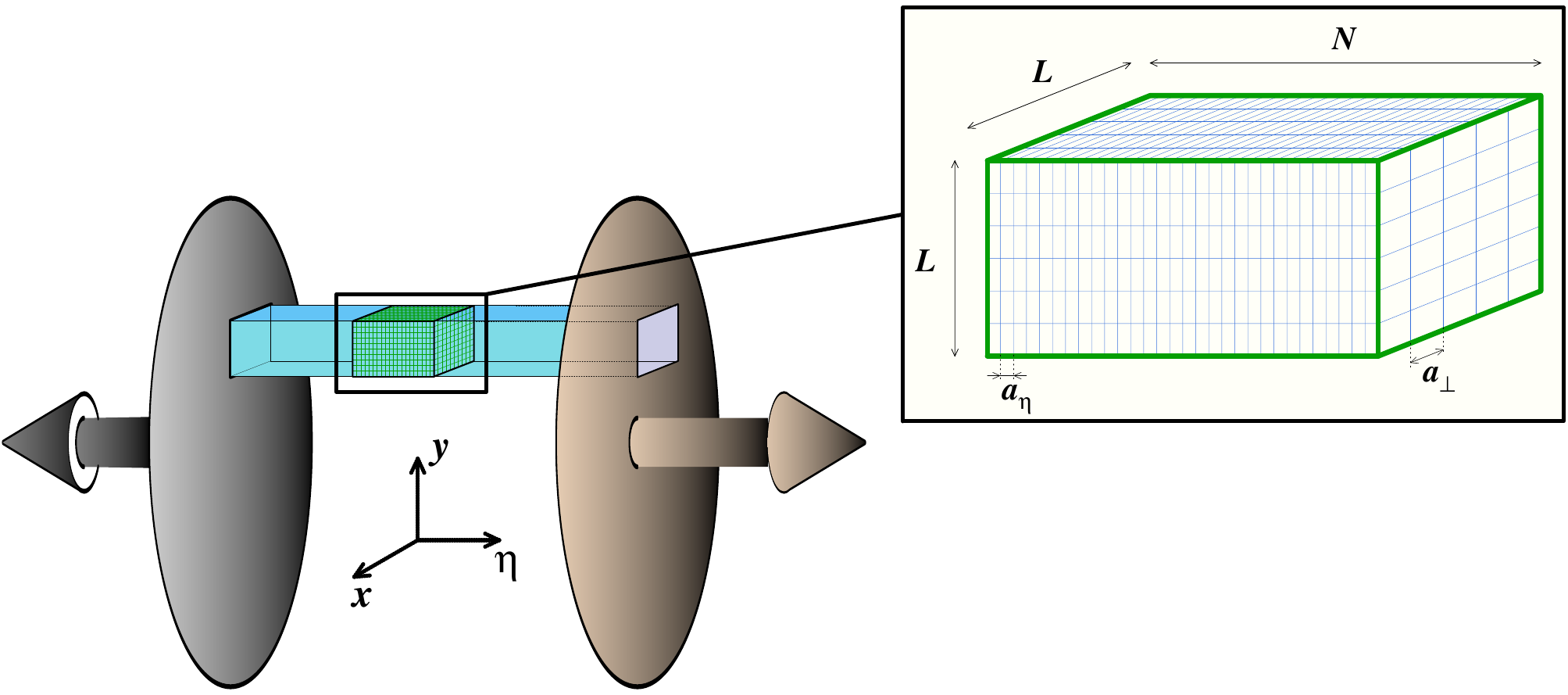}}
\hfill
\resizebox*{4cm}{!}{\includegraphics{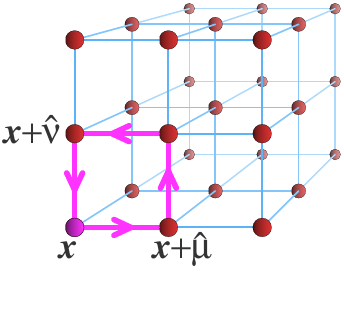}}
\end{center}
\caption{\label{fig:simul}Lattice setup for the numerical solution of the classical Yang-Mills equations.}
\end{figure}

In order to solve numerically the classical Yang-Mills equations, one
must discretize space. Moreover, one should use the
$(\x_\perp,\eta,\tau)$ system of coordinates, for the boost invariance
of a high energy collision to become manifest. Due to limited
computational resources, one usually simulates only a small part of the
collision domain, as shown in the left figure \ref{fig:simul}.
The plots that will be presented later in this section have been
obtained on a $64\times 64\times 128$ lattice (with a longitudinal
lattice spacing $a_\eta=1/64$, and a transverse lattice spacing $Q_s
a_\perp =1$)~\cite{EpelbG3}. In order to preserve on the lattice an
exact invariance with respect to time independent gauge
transformations, the gauge potentials must be traded in favor of
Wilson lines (link variables) that live on the edges of the lattice,
as shown in the right figure \ref{fig:simul}. The electrical fields,
that transform covariantly under these gauge transformations, live on
the nodes of the lattice. In this representation, the classical
Hamiltonian is a sum of the squares of the electrical fields, and of
the traces of the plaquettes (product of link variables along a closed
loop) that span the elementary squares of the lattice (see the figure
\ref{fig:simul}).

The equations of motion are solved with the leapfrog algorithm, with
time steps that are adjusted in order to conserve Gauss's law to very
good accuracy. From the solution of the Yang-Mills equations, we
compute the components of the energy-momentum tensor. By construction,
we have
\begin{equation}
2P_{_T}+P_{_L}=\epsilon\; .
\label{eq:tmumu}
\end{equation}
The energy density and longitudinal pressure are also related by
Bjorken's law,
\begin{equation}
\frac{\partial\epsilon}{\partial\tau}+\frac{\epsilon+P_{_L}}{\tau}=0\; ,
\label{eq:BL}
\end{equation}
which arises as a consequence of local energy and momentum conservation
for a boost invariant system.

A crucial step in this numerical calculation is the subtraction of the
terms that would be singular in the continuum limit (i.e. in the
ultraviolet). The most singular terms come from the zero point energy,
that produce a quartic divergence in the energy-momentum tensor. These
terms do not depend on the background, and can be removed by
performing the calculation a second time with $Q_s\equiv 0$. However,
this procedure is made difficult by the fact that the ultraviolet cutoff
on longitudinal momentum $p_z$ is time dependent ($p_z^{\rm max}\sim
a_\eta/\tau$) on a lattice with a fixed spacing in the rapidity
$\eta$. Therefore, the terms that should be subtracted become very
large at small $\tau$. Since their accuracy is limited by the
statistics used in the Monte-Carlo average over the initial
field fluctuations, the statistical errors in the difference are large at
small times.

After the zero point energy has been subtracted, the energy-momentum
tensor (a dimension 4 operator) can also mix with dimension 2
operators to produce a weaker quadratic ultraviolet divergence. In the
continuum, there is no local gauge invariant dimension 2 operators in
Yang-Mills theory, with which it could mix. At finite lattice spacing
however, an operator like the trace of a plaquette could play this
role, and produce terms that are quadratic in the inverse longitudinal
lattice spacing (they would affect mostly the small time behavior
since the longitudinal ultraviolet cutoff behaves as $1/\tau$). In
fact, such terms are seen numerically, in energy density and in the
longitudinal pressure (but not in the transverse pressure, which
indicates that they affect only the transverse chromo-electric and
chromo-magnetic fields). In order to preserve eq.~(\ref{eq:tmumu}),
the counterterms in $\epsilon$ and in $P_{_L}$ must be the same, and
Bjorken's law then tells us that this common counterterm must be of
the form $A/\tau^2$.  A deeper understanding would necessary in order
to compute the prefactor $A$ from first principles, and at the moment
it is simply fitted in order to eliminate the singular $1/\tau^2$
behavior in $\epsilon$ and in $P_{_L}$. The following equations
summarize the subtraction procedure~:
\begin{equation}
\begin{aligned}
\left<P_{_T}\right>_{{\rm phys.}}
&=
\left<P_{_T}\right>_{{\rm backgd.}\atop{\rm+\ fluct.}}
&-
\left<P_{_T}\right>_{{\rm fluct.}\atop{\rm only}}
&
\\
\left<\epsilon,P_{_L}\right>_{{\rm phys.}}
&
=
\underbrace{\left<\epsilon,P_{_L}\right>_{{\rm backgd.}\atop{\rm+\ fluct.}}}_{\rm computed}
&-
\underbrace{\left<\epsilon,P_{_L}\right>_{{\rm fluct.}\atop{\rm only}}}_{\rm computed}
&+
\underbrace{A\,\tau^{-2}\vphantom{\left<\epsilon,P_{_L}\right>_{{\rm fluct.}\atop{\rm only}}}}_{\rm fitted}
\; .
\end{aligned}
\end{equation}

\begin{figure}[htbp]
\begin{center}
\resizebox*{6.2cm}{!}{\includegraphics{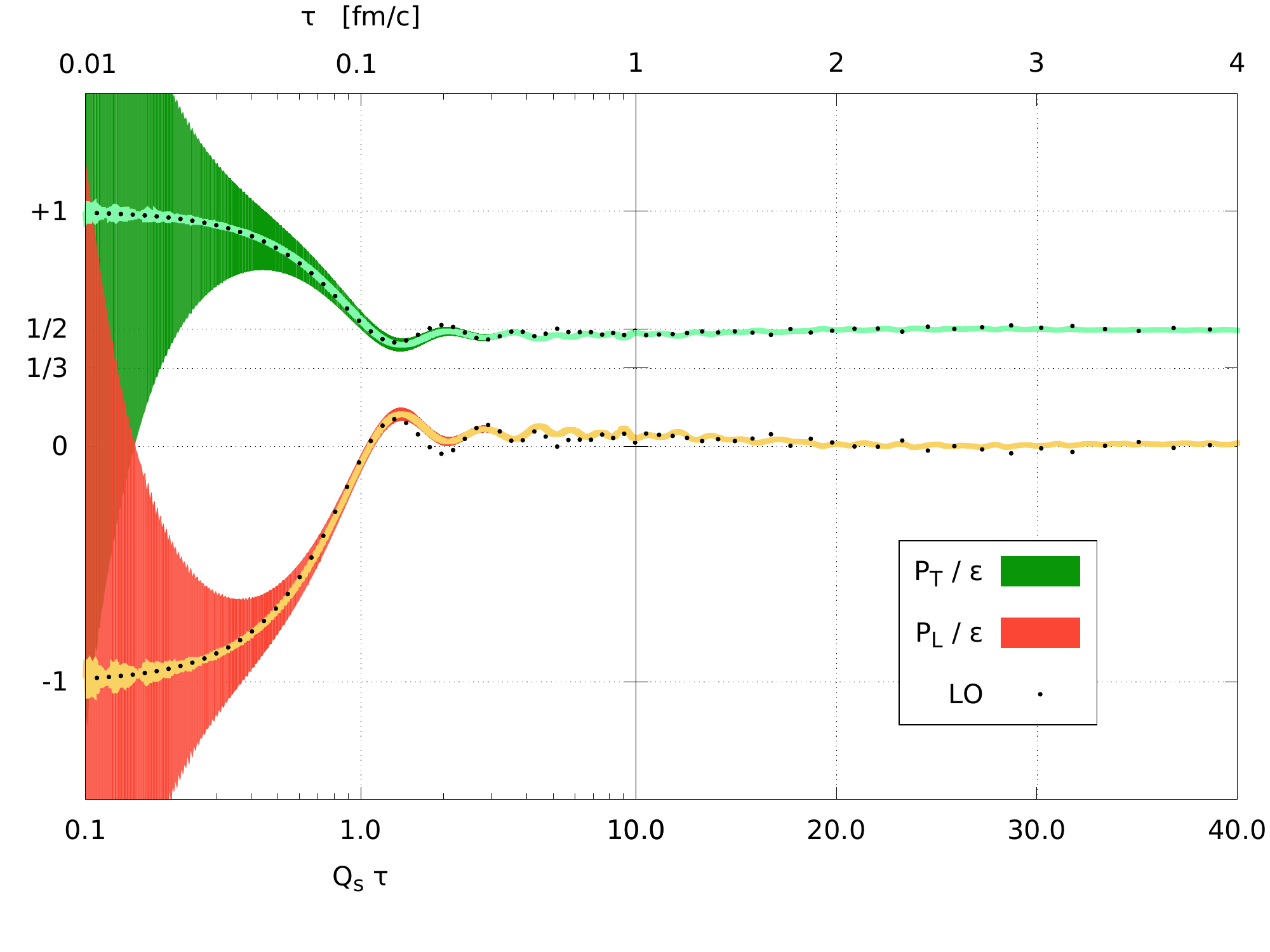}}
\hfill
\resizebox*{6.2cm}{!}{\includegraphics{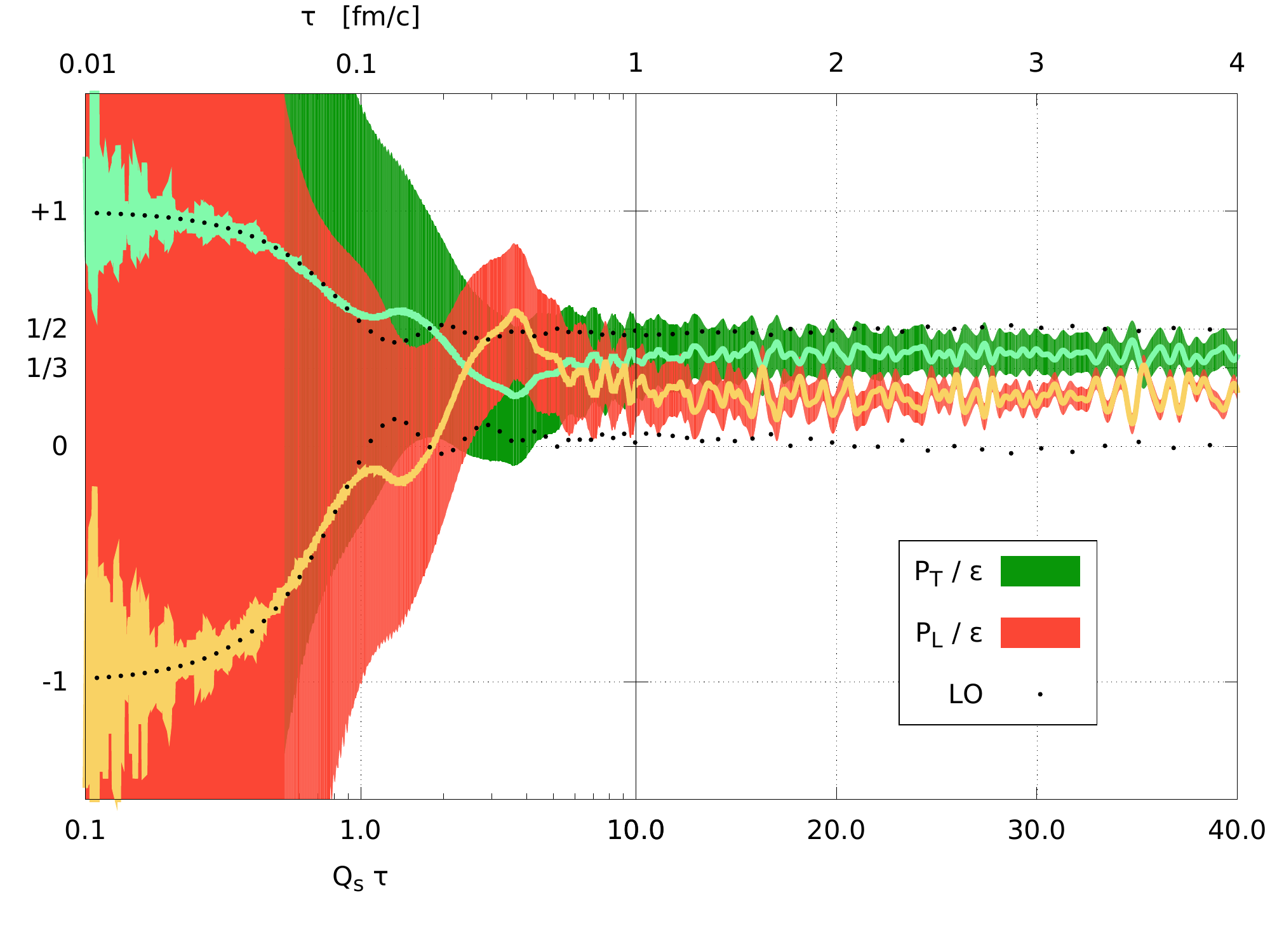}}
\end{center}
\caption{\label{fig:iso-YM}Time evolution of the ratios
  $P_{_T}/\epsilon$ and $P_{_L}/\epsilon$, for $g=0.1$ (left) and
  $g=0.5$ (right). The lower horizontal axis is in units of $Q_s\tau$,
  and the upper horizontal axis is in fm/c, assuming the value
  $Q_s=2~$GeV. The LO results are also shown for comparison.}
\end{figure}

Numerical results\footnote{A closely related study has also been performed
  in Refs.~\cite{BergeBSV1,BergeBSV2}. In that work, weaker couplings,
  later times (in units of $Q_s^{-1}$), and a completely incoherent
  distribution of initial fields were considered.} are displayed in
the figure \ref{fig:iso-YM}, for two values of the strong coupling
constant~: $g=0.1$ (left) and $g=0.5$ (right). The curves show the
time evolution of the ratios $P_{_L}/\epsilon$ and $P_{_T}/\epsilon$,
and the bands are a rough estimate of the statistical errors (they are
very large at small times, because one needs to subtract two
quantities varying as $1/\tau^2$, in order to get a result of order
$\tau^0$).
The lower horizontal axis is in units of $Q_s\tau$, while the upper
horizontal axis is in fm/c (the calibration of the time axis in
physical units requires that one chooses the value of $Q_s$, here
taken to be $Q_s=2~$GeV, a typical value for heavy ion collisions at
LHC energy). The dotted lines show the LO results for these two
ratios. As said before, there is no significant buildup of
longitudinal pressure at LO, and it remains much smaller than the
transverse pressure at all times. The first thing to note is that at
early times, $Q_s\tau\ll 1$, the resummed results are identical to the
LO results. In fact, since the Weibel instabilities need a time
$Q_s\tau\sim 1$ to develop, this was to be expected for the small
values of the coupling considered here (note that this agreement is
somewhat non-trivial because it relies on the proper subtraction of
the ultraviolet contributions). At the smallest of the two couplings,
the resummed result stays very close to the LO result at all times,
and there is no significant increase of the longitudinal pressure
compared to LO. At $g=0.5$ (which corresponds to a 25-fold increase of
$\alpha_s$), some qualitative changes in the behavior of the
longitudinal pressure become visible~: at times $Q_s\tau\gtrsim 2$ it
deviates significantly from the LO result and becomes comparable to
the transverse pressure. Concomitantly, the ratio $P_{_T}/\epsilon$
decreases from $1/2$ to a value which is much closer to $1/3$.  This
result suggests, that even at rather weak couplings, the higher order
corrections --that involve the modes that are subject to the Weibel
instability-- play an important role in the isotropization of the
pressure tensor.

\section{Summary}
In this short review, we have summarized our present understanding on
how to describe the early stages of high energy heavy ion collisions
in terms of quantum chromodynamics. 

A central tool for these studies is the Color Glass Condensate, an
effective theory for strong interactions in the regime of large gluon
densities. Indeed, the gluon distribution in a hadron increases with
energy due to repeated gluon emissions by bremsstrahlung. Eventually,
the gluon occupation number becomes of order $1/g^2$, and the
non-linear interactions among the gluons in the hadronic wavefunctions
can no longer be neglected -- a regime known as gluon saturation. The
CGC provides a framework to study the approach to saturation, and the
subsequent evolution of the gluon distribution into the saturation
regime.

The CGC is also the framework of choice in order to consistently
calculate the expectation value of observables after the collision of
two saturated projectiles. A consequence of having large gluon
occupation numbers in the two projectiles is that an infinite set of
graphs contribute at each order in $g^2$. The leading order is given
by the sum of all the tree graphs, which can be conveniently expressed
in terms of solutions to the classical Yang-Mills equations. The
next-to-leading order is the sum of all the 1-loop graphs, where the
loop is dressed by the classical field found at leading order. 

In addition to large logarithms of the collision energy (that can be
absorbed into the JIMWLK evolution of the distribution of color
sources), the NLO also contains secular terms, that grow exponentially
with time because of the Weibel instability. It is possible to collect
at each loop order the terms that have the fastest growth, and to
resum them. It turns out that these contributions can all be
attributed to fluctuations of the initial fields, whose time evolution
remains classical. A practical way of performing this resummation is
to repeat the same classical computation as the one needed at LO, with
a Gaussian distribution of initial fields which is sampled with a
Monte-Carlo. The variance of this Gaussian distribution has been
derived from a 1-loop calculation in \cite{EpelbG2}, which paves the
way for a numerical implementation of this resummation. The first
numerical results implementing this program have been obtained in
\cite{EpelbG3}, and show that the higher order terms included via
this resummation lead to an increase of the longitudinal pressure.

\vglue 2mm
\noindent{\bf Acknowledgements~:}
This work is support by the Agence Nationale de la Recherche project
11-BS04-015-01.


\begin{thebibliography}{10}

\bibitem{KarscC1}
{F. Karsch, et al. [RBC and HotQCD Collaboration]}, J. Phys. {\bf G35}, 104096
  (2008).

\bibitem{Adamsa3}
{J. Adams, et al.}, [STAR Collaboration] Nucl. Phys. {\bf A} {\bf 757}, 102
  (2005).

\bibitem{Adcoxa1}
{K. Adcox, et al.}, [PHENIX Collaboration] Nucl. Phys. {\bf A} {\bf 757}, 184
  (2005).

\bibitem{Arsena2}
{I. Arsene, et al.}, [BRAHMS collaboration] Nucl. Phys. {\bf A} {\bf 757}, 1
  (2005).

\bibitem{Backa2}
{B.B. Back, et al.}, [PHOBOS collaboration] Nucl. Phys. {\bf A} {\bf 757}, 28
  (2005).

\bibitem{HuoviR1}
{P. Huovinen, P.V. Ruuskanen}, Ann. Rev. Nucl. Part. Sci. {\bf 56}, 163 (2006).

\bibitem{Romat1}
{P. Romatschke}, Int. J. Mod. Phys. E {\bf 19}, 1 (2010).

\bibitem{Teane1}
{D. Teaney}, Prog. Part. Nucl. Phys. {\bf 62}, 451 (2009).

\bibitem{RomatR1}
{P. Romatschke, U. Romatschke}, Phys. Rev. Lett. {\bf 99}, 172301 (2007).

\bibitem{Aarona2}
{F.D. Aaron, et al}, [H1 and ZEUS Collaborations] JHEP {\bf 1001}, 109 (2010).

\bibitem{DeshpEM1}
{A. Deshpande, R. Ent, R. Milner}, CERN Courier, October 2009.

\bibitem{GriboLR1}
{L.V. Gribov, E.M. Levin, M.G. Ryskin}, Phys. Rept. {\bf 100}, 1 (1983).

\bibitem{MuellQ1}
{A.H. Mueller, J-W. Qiu}, Nucl. Phys. {\bf B} {\bf 268}, 427 (1986).

\bibitem{BlaizM2}
{J.P. Blaizot, A.H. Mueller}, Nucl. Phys. {\bf B} {\bf 289}, 847 (1987).

\bibitem{McLerV1}
{L.D. McLerran, R. Venugopalan}, Phys. Rev. {\bf D} {\bf 49}, 2233 (1994).

\bibitem{McLerV2}
{L.D. McLerran, R. Venugopalan}, Phys. Rev. {\bf D} {\bf 49}, 3352 (1994).

\bibitem{McLerV3}
{L.D. McLerran, R. Venugopalan}, Phys. Rev. {\bf D} {\bf 50}, 2225 (1994).

\bibitem{Schwi1}
{J. Schwinger}, J. Math. Phys. {\bf 2}, 407 (1961).

\bibitem{Keldy1}
{L.V. Keldysh}, Sov. Phys. JETP {\bf 20}, 1018 (1964).

\bibitem{Cutko1}
{R.E. Cutkosky}, J. Math. Phys. {\bf 1}, 429 (1960).

\bibitem{t'HooV1}
{G. t'Hooft, M.J.G. Veltman}, CERN report 73-9.

\bibitem{GelisV2}
{F. Gelis, R. Venugopalan}, Nucl. Phys. {\bf A} {\bf 776}, 135 (2006).

\bibitem{GelisV3}
{F. Gelis, R. Venugopalan}, Nucl. Phys. {\bf A} {\bf 779}, 177 (2006).

\bibitem{Kovch1}
{Yu.V. Kovchegov}, Phys. Rev. {\bf D} {\bf 54}, 5463 (1996).

\bibitem{KovneMW2}
{A. Kovner, L.D. McLerran, H. Weigert}, Phys. Rev. {\bf D} {\bf 52}, 6231
  (1995).

\bibitem{KrasnV3}
{A. Krasnitz, R. Venugopalan}, Nucl. Phys. {\bf B} {\bf 557}, 237 (1999).

\bibitem{KrasnV1}
{A. Krasnitz, R. Venugopalan}, Phys. Rev. Lett. {\bf 84}, 4309 (2000).

\bibitem{KrasnV2}
{A. Krasnitz, R. Venugopalan}, Phys. Rev. Lett. {\bf 86}, 1717 (2001).

\bibitem{KrasnNV2}
{A. Krasnitz, Y. Nara, R. Venugopalan}, Phys. Rev. Lett. {\bf 87}, 192302
  (2001).

\bibitem{KrasnNV1}
{A. Krasnitz, Y. Nara, R. Venugopalan}, Nucl. Phys. {\bf A} {\bf 727}, 427
  (2003).

\bibitem{Lappi1}
{T. Lappi}, Phys. Rev. {\bf C} {\bf 67}, 054903 (2003).

\bibitem{Lappi3}
{T. Lappi}, Phys. Lett. {\bf B} {\bf 643}, 11 (2006).

\bibitem{KrasnNV3}
{A. Krasnitz, Y. Nara, R. Venugopalan}, Phys. Lett. {\bf B} {\bf 554}, 21
  (2003).

\bibitem{KrasnNV4}
{A. Krasnitz, Y. Nara, R. Venugopalan}, Nucl. Phys. {\bf A} {\bf 717}, 268
  (2003).

\bibitem{LappiV1}
{T. Lappi, R. Venugopalan}, Phys. Rev. {\bf C} {\bf 74}, 054905 (2006).

\bibitem{LappiM1}
{T. Lappi, L.D. McLerran}, Nucl. Phys. {\bf A} {\bf 772}, 200 (2006).

\bibitem{FukusG1}
{K. Fukushima, F. Gelis}, Nucl. Phys. {\bf A} {\bf 874}, 108 (2012).

\bibitem{AyalaJMV2}
{A. Ayala, J. Jalilian-Marian, L.D. McLerran, R. Venugopalan}, Phys. Rev. {\bf
  D} {\bf 53}, 458 (1996).

\bibitem{IancuV1}
{E. Iancu, R. Venugopalan}, Quark Gluon Plasma 3, Eds. R.C. Hwa and X.N. Wang,
  World Scientific, hep-ph/0303204.

\bibitem{Lappi6}
{T. Lappi}, Int. J. Mod. Phys. {\bf E20}, 1 (2011).

\bibitem{Weige2}
{H. Weigert}, Prog. Part. Nucl. Phys. {\bf 55}, 461 (2005).

\bibitem{GelisIJV1}
{F. Gelis, E. Iancu, J. Jalilian-Marian, R. Venugopalan}, Ann. Rev. Part. Nucl.
  Sci. {\bf 60}, 463 (2010).

\bibitem{GelisLV3}
{F. Gelis, T. Lappi, R. Venugopalan}, Phys. Rev. {\bf D} {\bf 78}, 054019
  (2008).

\bibitem{GelisLV4}
{F. Gelis, T. Lappi, R. Venugopalan}, Phys. Rev. {\bf D} {\bf 78}, 054020
  (2008).

\bibitem{GelisLV5}
{F. Gelis, T. Lappi, R. Venugopalan}, Phys. Rev. {\bf D} {\bf 79}, 094017
  (2009).

\bibitem{Balit1}
{I. Balitsky}, Nucl. Phys. {\bf B} {\bf 463}, 99 (1996).

\bibitem{JalilKMW1}
{J. Jalilian-Marian, A. Kovner, L.D. McLerran, H. Weigert}, Phys. Rev. {\bf D}
  {\bf 55}, 5414 (1997).

\bibitem{JalilKLW1}
{J. Jalilian-Marian, A. Kovner, A. Leonidov, H. Weigert}, Nucl. Phys. {\bf B}
  {\bf 504}, 415 (1997).

\bibitem{JalilKLW2}
{J. Jalilian-Marian, A. Kovner, A. Leonidov, H. Weigert}, Phys. Rev. {\bf D}
  {\bf 59}, 014014 (1998).

\bibitem{JalilKLW3}
{J. Jalilian-Marian, A. Kovner, A. Leonidov, H. Weigert}, Phys. Rev. {\bf D}
  {\bf 59}, 034007 (1999).

\bibitem{JalilKLW4}
{J. Jalilian-Marian, A. Kovner, A. Leonidov, H. Weigert}, Phys. Rev. {\bf D}
  {\bf 59}, 099903 (1999).

\bibitem{IancuLM1}
{E. Iancu, A. Leonidov, L.D. McLerran}, Nucl. Phys. {\bf A} {\bf 692}, 583
  (2001).

\bibitem{IancuLM2}
{E. Iancu, A. Leonidov, L.D. McLerran}, Phys. Lett. {\bf B} {\bf 510}, 133
  (2001).

\bibitem{FerreILM1}
{E. Ferreiro, E. Iancu, A. Leonidov, L.D. McLerran}, Nucl. Phys. {\bf A} {\bf
  703}, 489 (2002).

\bibitem{RomatV1}
{P. Romatschke, R. Venugopalan}, Phys. Rev. Lett. {\bf 96}, 062302 (2006).

\bibitem{RomatV2}
{P. Romatschke, R. Venugopalan}, Eur. Phys. J. {\bf A} {\bf 29}, 71 (2006).

\bibitem{RomatV3}
{P. Romatschke, R. Venugopalan}, Phys. Rev. {\bf D} {\bf 74}, 045011 (2006).

\bibitem{BiroGMT1}
{T.S. Biro, C. Gong, B. Muller, A. Trayanov}, Int. J. Mod. Phys. {\bf C} {\bf
  5}, 113 (1994).

\bibitem{HeinzHLMM1}
{U.W. Heinz, C.R. Hu, S. Leupold, S.G. Matinyan, B. Muller}, Phys. Rev. {\bf D}
  {\bf 55}, 2464 (1997).

\bibitem{BolteMS1}
{J. Bolte, B. M{\accent "7F u}ller, A. Sch{\accent "7F a}fer}, Phys. Rev. {\bf
  D} {\bf 61}, 054506 (2000).

\bibitem{FujiiI1}
{H. Fujii, K. Itakura}, Nucl. Phys. {\bf A} {\bf 809}, 88 (2008).

\bibitem{FujiiII1}
{H. Fujii, K. Itakura, A. Iwazaki}, Nucl. Phys. {\bf A} {\bf 828}, 178 (2009).

\bibitem{KunihMOST1}
{T. Kunihiro, B. Muller, A. Ohnishi, A. Schafer, T.T. Takahashi, {\bf A}
  Yamamoto}, Phys. Rev. {\bf D} {\bf 82}, 114015 (2010).

\bibitem{Mrowc3}
{S. Mrowczynski}, Phys. Lett. {\bf {\bf B} 314}, 118 (1993).

\bibitem{Mrowc4}
{S. Mrowczynski}, Phys. Lett. {\bf {\bf B} 393}, 26 (1997).

\bibitem{RebhaRS1}
{A.K. Rebhan, P. Romatschke, M. Strickland}, Phys. Rev. Lett. {\bf 94}, 102303
  (2005).

\bibitem{RebhaRS2}
{A.K. Rebhan, P. Romatschke, M. Strickland}, JHEP {\bf 0509}, 041 (2005).

\bibitem{MrowcRS1}
{S. Mrowczynski, A. Rebhan, M. Strickland}, Phys. Rev. {\bf D} {\bf 70}, 025004
  (2004).

\bibitem{RomatS1}
{P. Romatschke, M. Strickland}, Phys. Rev. {\bf D} {\bf 68}, 036004 (2003).

\bibitem{RomatS2}
{P. Romatschke, M. Strickland}, Phys. Rev. {\bf D} {\bf 70}, 116006 (2004).

\bibitem{RebhaS1}
{A.K. Rebhan, D. Steineder}, Phys. Rev. {\bf D} {\bf 81}, 085044 (2010).

\bibitem{RebhaSA1}
{A.K. Rebhan, M. Strickland, M. Attems}, Phys. Rev. {\bf D} {\bf 78}, 045023
  (2008).

\bibitem{ArnolLM1}
{P. Arnold, J. Lenaghan, G.D. Moore}, JHEP {\bf 0308}, 002 (2003).

\bibitem{ArnolLMY1}
{P. Arnold, J. Lenaghan, G.D. Moore, L.G. Yaffe}, Phys. Rev. Lett. {\bf 94},
  072302 (2005).

\bibitem{ArnolM1}
{P. Arnold, G.D. Moore}, Phys. Rev. {\bf D} {\bf 73}, 025013 (2006).

\bibitem{ArnolM2}
{P. Arnold, G.D. Moore}, Phys. Rev. {\bf D} {\bf 73}, 025006 (2006).

\bibitem{ArnolM3}
{P. Arnold, G.D. Moore}, Phys. Rev. {\bf D} {\bf 76}, 045009 (2007).

\bibitem{ArnolMY4}
{P. Arnold, G.D. Moore, L.G. Yaffe}, Phys. Rev. {\bf D} {\bf 72}, 054003
  (2005).

\bibitem{DumitNS1}
{A. Dumitru, Y. Nara, M. Strickland}, Phys. Rev. {\bf D} {\bf 75}, 025016,
  (2007).

\bibitem{BodekR1}
{D. Bodeker, K. Rummukainen}, JHEP {\bf 0707}, 022 (2007).

\bibitem{BergeGSS1}
{J. Berges, D. Gelfand, S. Scheffler, D. Sexty}, Phys. Lett. {\bf B} {\bf 677},
  210 (2009).

\bibitem{BergeSS2}
{J. Berges, S. Scheffler, D. Sexty}, Phys. Rev. {\bf D} {\bf 77}, 034504
  (2008).

\bibitem{KurkeM1}
{A. Kurkela, G.D. Moore}, JHEP {\bf 1112}, 044 (2011).

\bibitem{KurkeM2}
{A. Kurkela, G.D. Moore}, JHEP {\bf 1111}, 120 (2011).

\bibitem{AttemRS1}
{M. Attems, A. Rebhan, M. Strickland}, Phys. Rev. {\bf D} {\bf 87}, 025010
  (2013).

\bibitem{DusliEGV1}
{K. Dusling, T. Epelbaum, F. Gelis, R. Venugopalan}, Nucl. Phys. {\bf A} {\bf
  850}, 69 (2011).

\bibitem{PolarS1}
{D. Polarski, A.A. Starobinsky}, Class. Quant. Grav. {\bf 13}, 377 (1996).

\bibitem{Son1}
{D.T. Son}, hep-ph/9601377.

\bibitem{KhlebT1}
{S.Yu. Khlebnikov, I.I. Tkachev}, Phys. Rev. Lett. {\bf 77}, 219 (1996).

\bibitem{MichaT1}
{R. Micha, I.I. Tkachev}, Phys. Rev. {\bf D} {\bf 70}, 043538 (2004).

\bibitem{FukusGM1}
{K. Fukushima, F. Gelis, L. McLerran}, Nucl. Phys. {\bf A} {\bf 786}, 107
  (2007).

\bibitem{EpelbG2}
{T. Epelbaum, F. Gelis}, Phys. Rev. {\bf D} {\bf 88}, 085015 (2013).

\bibitem{EpelbG3}
{T. Epelbaum, F. Gelis}, Phys. Rev. Lett. {\bf 111}, 232301 (2013).

\bibitem{BergeBSV1}
{J. Berges, K. Boguslavski, S. Schlichting, R. Venugopalan}, arXiv:1303.5650.

\bibitem{BergeBSV2}
{J. Berges, K. Boguslavski, S. Schlichting, R. Venugopalan}, arXiv:1311.3005.

\end{thebibliography}

\end{document}